\documentclass[useAMS,fleqn,usenatbib]{mn2e}
\usepackage{times}
\usepackage{amsmath}
\usepackage{bm}
\usepackage{url}
\usepackage{graphicx}
\usepackage{hyperref}

\input amssym.tex

\mathchardef\mhyphen="2D

\newcommand{\bq}{\textbf{q}}
\newcommand{\bk}{\textbf{k}}
\newcommand{\bx}{\textbf{x}}

\newcommand{\avg}[1]{\left\langle{#1}\right\rangle}

\newcommand{\hmpc}{\,$h^{-1}$\,Mpc}

\newcommand{\hmpcnosp}{$h^{-1}$\,Mpc}

\newcommand{\LCDM}{$\Lambda$CDM}

\newcommand{\eg}{e.g., }

\newcommand{\bnabla}{\mbox{\boldmath $\nabla$}}
\newcommand{\bPsi}{\mbox{\boldmath $\Psi$}}

\newcommand{\divqpsi}{{\bnabla_q\cdot{\bPsi}}}
\newcommand{\divqx}{{\bnabla_q\cdot{\bx}}}
\newcommand{\dlin}{\delta_{\rm lin}}
\newcommand{\deltaL}{\delta_{\rm Lag}}
\newcommand{\deltaE}{\delta_{\rm Eul}}
\newcommand{\AL}{A_{\rm Lag}}
\newcommand{\AEul}{A_{\rm Eul}}
\newcommand{\psil}{\psi_{\rm lin}}
\newcommand{\psitwo}{\psi_{\rm 2LPT}}
\newcommand{\psiparab}{\psi_{\rm 2LPT, parab}}

\newcommand{\psisc}{\psi_{\rm sc}}
\newcommand{\psihalfexp}{\psi_{\rm halfexp}}
\newcommand{\org}{{\scshape origami}}

\chardef\til=`\~

\begin{document}

\title{Quantifying distortions of the Lagrangian dark-matter mesh in
  cosmology}

\author[Mark C.\ Neyrinck]
{Mark C.\ Neyrinck$^1$\\
$^{1}$Department of Physics and Astronomy, The Johns Hopkins University, Baltimore, MD 21218, USA}


\maketitle

\begin{abstract}
  We examine the Lagrangian divergence of the displacement field,
  arguably a more natural object than the density in a Lagrangian
  description of cosmological large-scale structure.  This quantity,
  which we denote $\psi$, quantifies the stretching and distortion of
  the initially homogeneous lattice of dark-matter particles in the
  universe.  $\psi$ encodes similar information as the density, but
  the correspondence has subtleties.  It corresponds better to the
  log-density $A$ than the overdensity $\delta$.  A Gaussian
  distribution in $\psi$ produces a distribution in $A$ with slight
  skewness; in $\delta$, we find that in many cases the skewness is
  further increased by 3.

  A local spherical-collapse-based (SC) fit found by Bernardeau gives
  a formula for $\psi$'s particle-by-particle behavior that works
  quite well, better than applying Lagrangian perturbation theory
  (LPT) at first or second (2LPT) order.  In 2LPT, there is a roughly
  parabolic relation between initial and final $\psi$ that can give
  overdensities in deep voids, so low-redshift, high-resolution 2LPT
  realizations should be used with caution.  The SC fit excels at
  predicting $\psi$ until streams cross; then, for particles forming
  haloes, $\psi$ plummets as in a waterfall to $-3$.  This gives a new
  method for producing $N$-particle realizations.  Compared to LPT
  realizations, such SC realizations give reduced stream-crossing, and
  better visual and 1-point-PDF correspondence to the results of full
  gravity.  LPT, on the other hand, predicts large-scale flows and the
  large-scale power-spectrum amplitude better, unless an empirical
  correction is added to the SC formula.
\end{abstract}

\begin {keywords}
  large-scale structure of Universe -- cosmology: theory
\end {keywords}

\section{Introduction}
In the present, quite observationally successful theory of cosmology,
the universe began with nearly uniform density everywhere.  According
to the theory of inflation, the small fluctuations in it began as tiny
quantum fluctuations, that `inflated' to macroscopic size as the
universe ballooned in its first instants.

In an Eulerian description, the density and velocity fields at fixed
comoving positions describe this process of structure formation.  In a
Lagrangian description, on the other hand, the fundamental object is
the displacement field, a vector field measuring the comoving distance
particles have traveled from their initial positions.  Fluctuations
are not fundamentally in the density, but in the separations between
particles.  If the particles are arranged on a cubic lattice, as they
often are in $N$-body simulations, the density fluctuations are
really deformations of this lattice.  In underdense regions, the
lattice stretches out; in overdense regions, it bunches together and
forms structures.

While the density is still relevant in a Lagrangian framework (as it
sources gravity), the simplest scalar to construct from the
displacement field (besides its magnitude, which is irrelevant for
local physics) is its Lagrangian divergence.  We denote this
divergence as $\psi$.  It is the lowest-order invariant (with respect
to rotations and translations) of the tidal tensor of the displacement
field, and quantifies the angle-averaged stretching of the Lagrangian
sheet in comoving coordinates.  Where a mass element becomes stretched
out, $\psi$ increases, and its density decreases.  For a potential
displacement field (i.e.\ with zero curl), $\psi$ carries all of its
information.

Lagrangian dynamics have long been applied to cosmology, going back at
least to \citet{Zeldovich1970}.  It can be insightful to envision the
process of structure formation in terms of the dynamics of a
Lagrangian `sheet,' a viewpoint that for instance has been applied to
classify the types of caustics (folds of this sheet) that can form
\citep[e.g.][]{ArnoldEtal1982,Arnold2001}.  This viewpoint has gotten
some attention recently
\citep[e.g.][]{ShandarinEtal2012,AbelEtal2011,FalckOrigami2012,Neyrinck2012}.
The `sheet' is initially flat in six-dimensional position-velocity
phase space, with vanishing bulk velocity everywhere as the cosmic
scale factor $a\to0$.  Seen in position space, gravity stretches out
the sheet in underdense regions, and bunches it together in overdense
regions.  Assuming cold dark matter, the sheet never intersects itself
in six-dimensional phase space, and instead folds up in rough analogy
to origami.

Several modifications to the original Zel'dovich approximation (ZA)
were proposed, including higher-order Lagrangian perturbation theory
(LPT).  In LPT, streams of matter typically overcross in high-density
regions, failing to form the bound structures that they would in full
gravity.  Several modifications of LPT \citep[e.g.][and references
  therein]{ColesEtal1993,MelottEtal1994} have been proposed to solve
this problem, another attempted solution for which also occurs in the
present paper.  The adhesion model, for instance
\citep{KofmanShandarin1988,KofmanEtal1992,Shandarin2009,ValageasBernardeau2011,HiddingEtal2012}
prevents stream crossing by introducing an effective viscosity,
allowing the structure to be predicted in an elegant geometrical
fashion.  There are other ways of producing approximate particle
realizations in a Lagrangian perturbation-theory approach, e.g.\ the
{\scshape pinocchio} algorithm \citep{MonacoEtal2002}.

One might argue that $N$-body simulations have become computationally
cheap enough that such approximate realizations have little use.  For
example, though, such techniques have proven quite useful in Bayesian
initial-conditions reconstruction
\citep{KitauraAngulo2012,JascheWandelt2012}, where full $N$-body
simulations would be far too slow.

In this paper, we explore the behavior and properties of the
Lagrangian spatial-stretching parameter $\psi$.  In Section
\ref{sec:approx}, we review several approximations for $\psi$ in the
literature and its relationship to the density field. In Section
\ref{sec:local}, we explore the relationship between $\psi$ and
density variables (the overdensity and the log-density) in a class of
`local Lagrangian' toy models, including a spherical collapse (SC)
approximation that is most relevant to structure formation.  In
Section \ref{sec:fullgravity}, we compare these approximations to
results from an $N$-body simulation, demonstrating the success of the
SC approximation.  In Section \ref{sec:realizations}, we test a simple
new way of producing particle realizations using the SC approximation,
and compare it to LPT approaches.

\section{Approximations for the displacement divergence}
\label{sec:approx}
There are several analytical approximations in the literature for the
Lagrangian divergence of the displacement field.  Where $\bq$ is a
Lagrangian coordinate of a particle, we denote the displacement field
as $\bPsi(\bq) = \bx(\bq)-\bq$ (with $\bx$ the Eulerian position of a
particle), and $\psi(\bq)\equiv\divqpsi(\bq)$.  Here
$(\bnabla_q\cdot)$ is the divergence operator in Lagrangian
coordinates.  Assuming that $\bPsi$ is a potential field,
$\psi(\bq)=\nabla^2_q\phi(\bq)$, where $\phi$ is the displacement
potential.  All of the approximations used in this paper assume that
$\bPsi$ is a potential field, which implies that $\psi$ contains all
of the information in $\bPsi$.  In this section, we examine a few of
these approximations.  In full gravity, $\Psi$ is not potential,
i.e.\ it has a nonzero curl.  However, as we show below, much of the
large-scale clustering is captured with the potential-flow assumption.

\subsection{Lagrangian Perturbation Theory}

The Zel'dovich approximation \citep[][ZA]{Zeldovich1970} is the
first-order, linear approximation in Lagrangian perturbation theory
(LPT).  The ZA gives
\begin{equation}
  \psil(\bq,\tau) = -\dlin(\bq,\tau)=-\frac{D_1(\tau)}{D_1(\tau_0)}\dlin(\bq,\tau_0),
  \label{eqn:zeld}
\end{equation}
where $\dlin$ is the overdensity linearly extrapolated with the linear
growth factor $D_1$, and $\tau_0$ is some initial time.

The second-order (2LPT) expression is more complicated, but still
straightforward to implement.  This slight added complexity seems
worth the trouble for initial-conditions generation
\citep{Scoccimarro1998,CrocceEtal2006,TatekawaMizuno2007,McCullaghInprep},
giving much-reduced `transients' compared to ZA-produced initial
conditions.

At second order,
\begin{equation}
  \psitwo(\bq) = \nabla^2_q\phi = -D_1\nabla^2_q\phi^{(1)}+D_2\nabla^2_q\phi^{(2)},
  \label{eqn:lpt21}
\end{equation}
where $\phi(\bq)$ is the total displacement potential, and $D_2$ is the
second-order growth factor. $D_2(\tau)\approx -\frac{3}{7} D_1^2(\tau)$, the
approximation holding to better than 2.6 percent for $0.1<\Omega_m<1$
\citep{BouchetEtal1995}.  These first- and second-order potentials are
\begin{align}
  \nabla^2_q\phi^{(1)}(\bq) & = \dlin(\bq),\\
  \nabla^2_q\phi^{(2)}(\bq) & = \sum_{i>j}\left\{\phi_{{\bf ,}ii}^{(1)}(\bq)\phi_{{\bf ,}jj}^{(1)}(\bq)-\left[\phi^{(1)}_{{\bf ,}ij}(\bq)\right]^2\right\}.
  \label{eqn:phi12}
\end{align}

We introduce here an isotropic, parabolic approximation to 2LPT, $\psiparab$, for $\psitwo$.   Note that (suppressing the $^{(1)}$ superscripts)
\begin{equation}
  \sum_{i>j}\left(\phi_{{\bf ,}ii}\phi_{{{\bf ,}jj}}\right) = \frac{1}{2}\left[\left(\nabla^2_q\phi \right)^2 - \sum_i {\phi_{{\bf ,}ii}}^2\right].
  \label{eqn:approx1}
\end{equation}
Also note that in 3D, $\sum_i {\phi_{{\bf ,}ii}}^2$ is bounded by $\frac{1}{3}(\nabla^2_q\phi)^2$ (in the isotropic case that $\phi_{{\bf ,}ii}$ are equal for all $i$), and $(\nabla^2_q\phi)^2$ (in the case that $\phi_{{\bf ,}ii} = \nabla^2_q\phi$ for some $i$, with all other components zero).  Putting this together, with $1/6$ (the isotropic case) $\ge c_2 \ge 1/2$, and recalling that $D_2<0$,
\begin{align}
  \psitwo(\bq) & = -D_1\delta + D_2\left[c_2\delta^2-\sum_{i>j}\left(\phi^{(1)}_{{\bf ,}ij}\right)^2\right]\\
  & \ge -D_1\delta + \frac{1}{3} D_2\delta^2\\
  & \approx -\dlin + \frac{1}{7}\dlin^2 \equiv \psiparab(\bq),
  \label{eqn:psiparab}
\end{align}
in the last line using the above approximation for $D_2$.

\begin{figure}
  \begin{center}
    \includegraphics[scale=0.4]{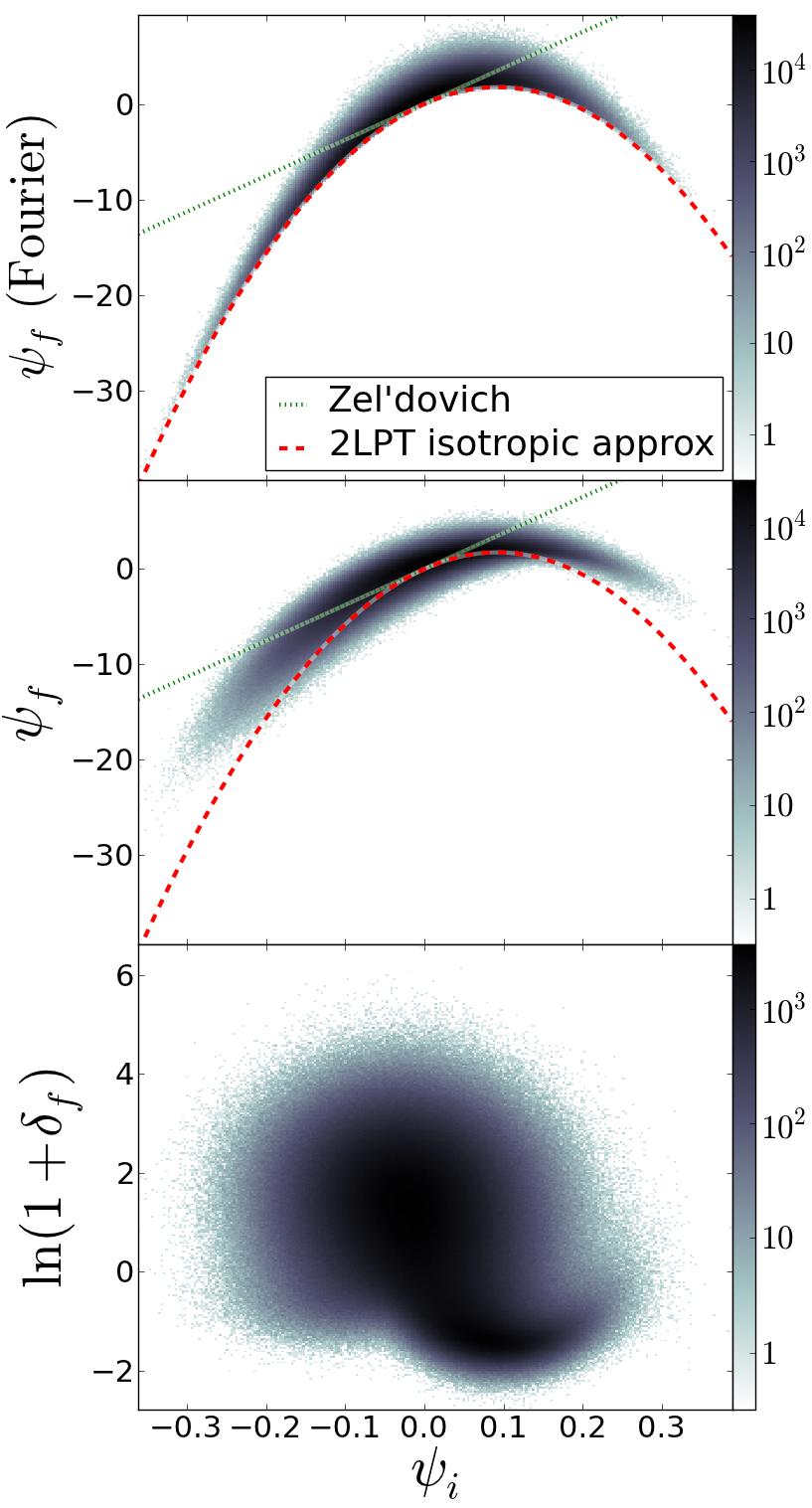}
  \end{center}  
  \caption{Two-dimensional histograms showing the relation between the
    stretching parameter $\psi(\bq)=\divqpsi(\bq)$, as well as the
    log-density, in the initial and final conditions, advancing
    particles according to second-order Lagrangian perturbation theory
    (2LPT).  In the top panel, $\psi$ is measured in Fourier space; in
    the middle, it is measured by differencing particle positions.
    The density in the bottom panel is estimated using a Voronoi
    tessellation, and is affected by multi-streaming.  The color of
    each grid cell corresponds to the number of particles (out of
    $256^3$) in that bin.  In the top panels, dotted green lines show
    the linear relationship in the Zel'dovich approximation, and the
    dashed red curve shows the isotropic approximation $\psiparab$ of
    Eq.\ \ref{eqn:psiparab}.  }
  \label{fig:densmap2lpt}
\end{figure}

Fig.\ \ref{fig:densmap2lpt} shows the 2LPT mapping between $\psi_i$
and $\psi_f$ ($\divqpsi$ at redshifts 49 and 0, respectively), as well
as the log-density $\ln(1+\delta)$, for particles in a
\LCDM\ $256^3$-particle, 200\hmpc-box size $N$-body simulation,
analyzed and discussed further below.  The particles were advanced
using the 2LPT algorithm as described by \citet{Scoccimarro1998}.  In
this standard technique, also used in producing a ZA realization,
first a Gaussian random field $\psi_i$ is generated. $\psi_f$ is then
estimated using an LPT approximation, and the final displacement field
is generated in Fourier space with an inverse-divergence operator.

For the top panel, the divergence was measured in Fourier space, the
native technique in the algorithm that produced the particle
distribution,
\begin{equation}
  (\bnabla\cdot\bpsi)_k=-i\bk\cdot\bpsi_k.
  \label{eqn:fourierdiv}
\end{equation}
For the middle panel, and in the rest of this paper, $\psi(\bq)$ was
measured in real space by differencing Eulerian positions of particles
immediately before and after the particle at position $\bq$, in
Lagrangian rows and columns of the initial lattice along the three
Cartesian directions.

There is a noticeable difference between the two methods of measuring
$\psi_f$ in the top panels.  One reason for this is that the
effective resolution of the real-space $\psi_f$ estimator is twice
that of the Fourier-space estimator.

Particularly using the Fourier-space estimator, the 2LPT prediction
fails at high $\psi$. Naively equating $\psi_f$ and $-\delta$, this
predicts strong overdensities in initial underdense regions!  This
behavior is tempered in the real-space-estimated $\psi_f$, but still
there are many apparently overdense, initially underdense particles.

This raises the question of whether artificial haloes might pop up in
what should be voids using 2LPT.  This is important since 2LPT is
sometimes used to generate low-redshift density distributions, for
example in modelling a sparsely sampled, large-volume survey, where
only the clustering on large (e.g.\ baryon-acoustic-oscillation)
scales needs to be accurate
\citep[e.g.][]{pthalos,NeyrinckSzapudi2008,ManeraEtal2012}.

As one test of this issue, we show the density as well in
Fig.\ \ref{fig:densmap2lpt}, measured with a Voronoi tessellation
\citep[\eg][]{svdw,voboz,vdws}.  For this Voronoi density estimate,
each particle occupies a Voronoi cell, a locus of points closer to
that particle than to any other particle.  The overdensity
$\delta_{\rm VTFE}=\avg{V}/V-1$ at a particle is set according to the
volume $V$ of its cell.  This density measure is mass-weighted, and
thus in a sense Lagrangian, but only strictly Lagrangian without
multi-streaming, which does occur in this 2LPT realization.

In the bottom panel of Fig.\ \ref{fig:densmap2lpt}, at moderate to
high densities, there is little correlation between $\psi_i$ and
$\delta$.  The over-shell-crossing in LPT, evident below in
Fig.\ \ref{fig:euler}, is one reason for this.  At low densities,
there are indeed a few overdense particles that have low $\psi_i$.  We
find this also below in Fig.\ \ref{fig:denscomp}, after which we further discuss this issue.

Previous authors
\citep{BuchertEtal1994,BouchetEtal1995,SahniShandarin1996} have noted
the failure of 2LPT in voids; they also found that going to
third-order LPT (3LPT) does not improve agreement substantially.  3LPT
comes at the expense of significantly greater complexity, and the
addition of a nonzero curl component (as exists in full gravity, as
well).  Since we have adopted the approximation that the displacement
field is potential in this paper, we stop our LPT analysis at second
order.

\subsection{The Spherical-Collapse Approximation}
\citet{Bernardeau1994} gave a simple formula for the evolution of an
average Lagrangian volume element, which gives a good fit to results
based on the spherical-collapse (SC) model.  It is based on an
$\Omega_M\to 0$ (and $\Omega_\Lambda=0$) limit he
\citep{Bernardeau1992} found to the nonlinear spherical-collapse
evolution of density.  A concise, instructive derivation appears in
\citep{BernardeauEtalReview2002}.  Since the formula arises from a
low-density limit, it is not surprising that it is quite accurate in
voids in $\Lambda$CDM, so perhaps we should call it the
spherical-expansion approximation.  In this approximation, the mass
element's volume, where $V_0$ is the mean volume occupied by a
particle (assuming equal masses), is
\begin{equation}
  V(t) = V_0\left(1-\frac{2}{3}\dlin\right)^{3/2}.
  \label{eqn:bsc}
\end{equation}
This approximation matches the behavior of `rare events' (matter in
deep voids) well.  \citet{FosalbaGaztanaga1998I} discuss how this
approximation arises, to leading order, in a monopole approximation to
Lagrangian perturbation theory.  This expression has proven to be
quite accurate, even outside of the `rare-event' limit in which it was
originally proposed, over a wide range of regimes and cosmologies
\citep{FosalbaGaztanaga1998III}.  \citet{ScherrerGaztanaga2001} found
that including a full parametric description of spherical-collapse
dynamics further improved matters, however at the expense of
additional complication and computation.

To get an equation for the time evolution of $\psi$ out of this, we
use a geometric isotropic-cube approximation to relate $\delta$ and
$\psi$.  The below derivation also essentially appears in
\citet{MohayaeeEtal2006}.  Assuming a Lagrangian mass element occupies
a cube of side length $\psi/3+1$ (giving $\divqpsi=\psi$), the volume
of such a cube in units of the mean volume is
\begin{equation}
  V=1/(1+\delta)=(1+\psi/3)^3.
  \label{eqn:volpsi}
\end{equation}

Equating the RHS of Eq.\ (\ref{eqn:volpsi}) to the volume in
Eq.\ \ref{eqn:bsc}, and employing the ZA $\psil=-\dlin$ gives what we
call the spherical-collapse (SC) approximation.  

\begin{align}
  \psisc & = 3\left[\left(1+\frac{2}{3}\psil\right)^{1/2}-1\right]\label{eqn:psisc}\\
  & = \psil-\frac{1}{6}\psil^2+\frac{1}{18}\psil^3+O(\psil^4),\nonumber
\end{align}
where $\psil=\frac{D_1(\tau)}{D_1(\tau_0)}\psi_0$ for some initial
time $\tau_0$.  This is one of a class of local approximations in
which $\psi_f$ at a given final redshift and position depends only on
its linear value, $\psil$.  In higher-order LPT, $\psi_f$ is generally
nonlocal, depending on derivatives of $\psil$, as well.

\section{Local Lagrangian approximations} 
\label{sec:local}
To explore some general relationships between the `stretching
parameter' $\psi$ and density variables, we further explore them in a
simple class of toy `local Lagrangian' models introduced by
\citet[][PS97]{ProtogerosScherrer1997}.  These models are
parameterized by $1<\alpha<3$,
\begin{equation}
  \delta_\alpha(\psi)=(1+\psi/\alpha)^{-\alpha}-1.
  \label{eqn:deltaalpha}
\end{equation}
Here, $\psi$ is the actual $\psi$ of a volume element, not necessarily
related to the linearly evolved $\psil$.

It may help to think of $\alpha$ conceptually as the effective number
of axes along which volume elements are expanding or contracting.  The
cubic-mass-element approximation in Eq.\ (\ref{eqn:volpsi}) has
$\alpha=3$.  However, confusingly, the $\alpha=3$ relationship was
used in deriving the $\alpha=3/2$ SC approximation.  $\alpha$ turns
from 3 to 3/2 only when we add the spherical-collapse relationship of
Eq.\ (\ref{eqn:bsc}), which relates (`final') $\psi$ to the linearly
evolved $\psil$.

The $\alpha=3/2$ model is particularly useful, but $\alpha$ may take
other effective values in other environments.  So, we do not confine
our attention exclusively to $\alpha=3/2$.

In these models, density singularities arise at $\psi=-\alpha$, where
the volume element has contracted to zero.  In the SC approximation
the critical density of a collapsed element is $-\psil=1.5$, close to
the Einstein-de Sitter linear spherical-collapse density, 1.69.

One way to quantify the non-linearity of the $\psi$-$\delta$
relationship is in its Taylor-series
coefficients. Eq.\ \ref{eqn:deltaalpha} expands to
\begin{equation}
  \delta_\alpha(\psi) = -\psi+\frac{1+\alpha}{2\alpha}\psi^2 -
  \frac{(1+\alpha)(2+\alpha)}{6\alpha^2}\psi^3+O(\psi^4).
  \label{eqn:deltacoeff}
\end{equation}

In this family of approximations, the log-density
\begin{equation}
  A_\alpha(\psi)\equiv\ln(1+\delta_\alpha)=-\alpha \ln(1+\psi/\alpha)
\end{equation}
has a much more linear relationship to $\psi$ than $\delta$ does
(recalling that $\alpha \ge 1$):
\begin{equation}
  A_\alpha(\psi)=-\psi+\frac{1}{2\alpha}\psi^2-\frac{1}{3\alpha^2}\psi^3+O(\psi^4).
  \label{eqn:acoeff}
\end{equation}
Curiously, the log-density is also closely related to the Eulerian
divergence of the displacement field\citep{FalckLogdens2012}, perhaps
even more so than the Lagrangian divergence, as we investigate here

In the next section, we will see that the distribution of $A$ given a
Gaussian-distributed $\psi$ is also significantly more Gaussian than
$\delta$.

\begin{figure}
  \begin{center}
    \includegraphics[scale=0.45]{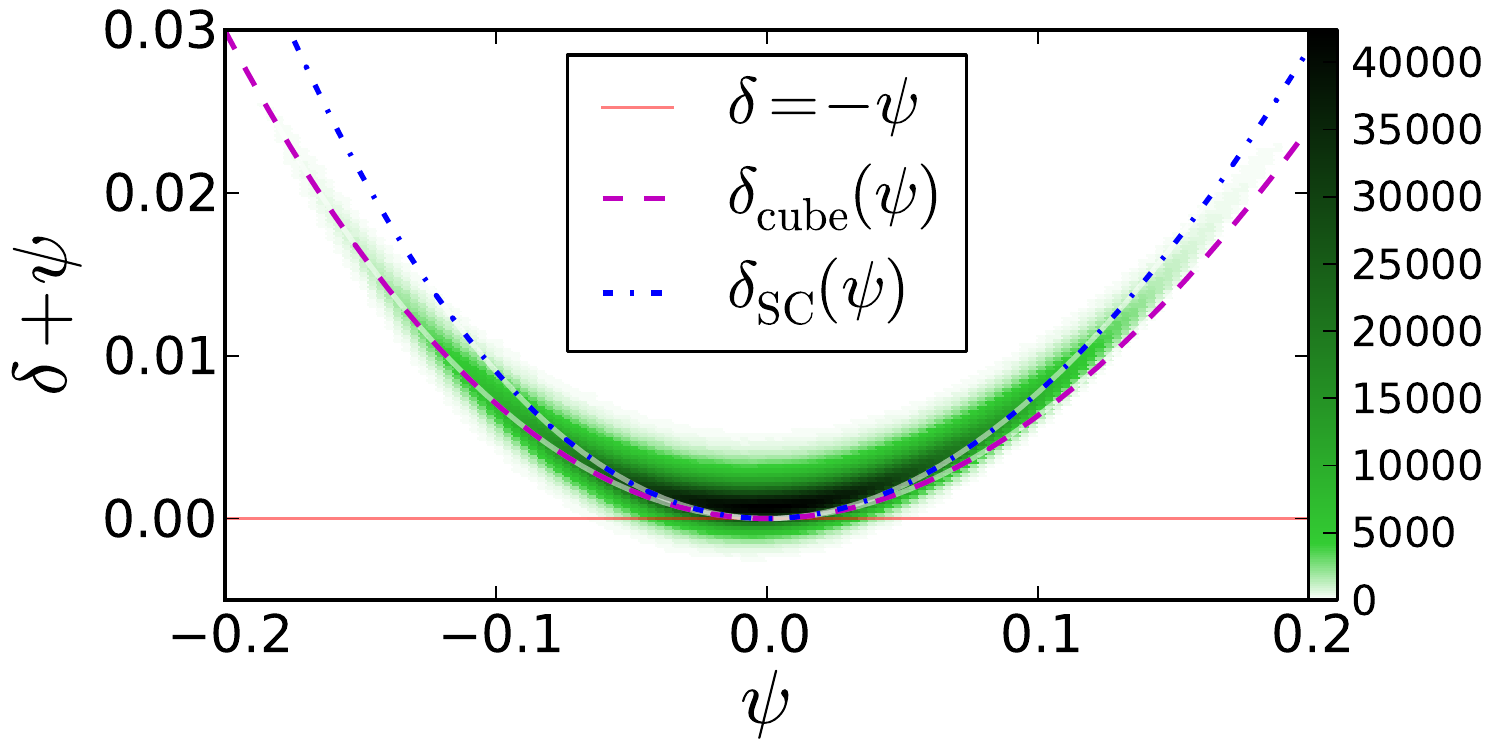}
  \end{center}  
  \caption{In green and black, two-dimensional histograms showing the
    relation between $\psi(\bq)=\divqpsi(\bq)$, and $\delta$, for
    $256^3$ particles in a set of $z=49$ initial conditions produced
    using the Zel'dovich approximation (ZA).  The deviation from
    $\psi=-\delta$ follows the cubic-mass-element approximation
    $\delta_{\rm cube}$ for $\psi<0$; for $\psi>0$, the result is
    between $\delta_{\rm cube}$ and $\delta_{\rm SC}$.  The histogram
    is unnormalized, showing the number of particles in each bin.}
  \label{fig:initvoldelta}
\end{figure}

At early epochs in the Zel'dovich approximation, a cubic-mass-element
model with $\alpha=3$ describes the density distribution quite well.
Fig.\ \ref{fig:initvoldelta} shows the accuracy of this
cubic-mass-element relationship is in a set of ZA-produced
\LCDM\ initial conditions at redshift $z=49$.  This simulation, used
below, has 256$^3$ particles, and box size 200\hmpc.  Again, the
density was estimated with a Voronoi method at each particle; here, it
is a true Lagrangian density estimate, since the fluctuations are
small enough that no multi-streaming has occurred.

Although the nonlinearity in Fig.\ \ref{fig:initvoldelta} is a bit
accentuated by the stretched $y$-axis, it is still substantial.
Putting $\alpha=3$ in Eqs.\ (\ref{eqn:deltacoeff}) and
(\ref{eqn:acoeff}), the $\psi^2$ coefficients in $\delta$ and $A$ are
$2/3$ (rather large; by far the highest among the approximations here
explored) and $1/6$. 

\subsection{Density PDFs}
\label{sec:pdfs}
Analytical density PDFs easily emerge from such local Lagrangian
approximations, which consist of simple transformations on initial
distributions.  Here we assume a Gaussian distribution in $\psi$, but
a non-Gaussian $\psi$ could also be transformed.

Note that even without explicit initial non-Gaussianity at arbitrarily
early times, a Gaussian distribution in $\psi$ results in a
non-Gaussian $\delta$ distribution.  If a Gaussian $\delta$
distribution is truly desired, one could start with an appropriately
non-Gaussian $\psi$ distribution, though we do not explore this
possibility here.

We transform the distributions with the change-of-variables formula
\begin{equation}
  P(y)=P(x)\left|dx/dy\right|,
  \label{eqn:changeofvars}
\end{equation}
where $P(x)dx$ and $P(y)dy$ give the PDFs of variables $x$ and $y$.

PS97 worked out the PDF of $\delta$ for the above
$\alpha$-parameterized local-Lagrangian approximations, allowing shell
crossing by using the absolute value of the volume element in
Eq.\ (\ref{eqn:deltaalpha}).  Here, we take a slightly different
approach, removing volume elements that have undergone shell crossing
(i.e.\ with $1+\psi/\alpha<0$) from consideration.  This leads to a
PDF that does not integrate entirely to 1, although its integral is
bounded below by 1/2 for large $\sigma_\psi$, and differs negligibly
from 1 for $\sigma_\psi\lesssim 0.5$.  Assuming the fraction of such
removed particles is small, i.e.\ for $\sigma_\psi\lesssim 0.5$, the
PDF of $\deltaL$, the mass-weighted density distribution, is

\begin{equation}
  P(\deltaL)=\frac{\exp\left[-\alpha^2\left\{(1+\delta)^{-1/\alpha}-1\right\}^2/(2\sigma_\psi^2)\right]}
  {(1+\deltaL)^{-1-1/\alpha}\sqrt{2\pi\sigma_\psi^2}}
  \label{eqn:pdfdeltaL}
\end{equation}
An Eulerian PDF for $\delta$ can be obtained by multiplying the
Lagrangian PDF by a factor of $V/\langle V\rangle$, where
$V=1/(1+\delta)$, giving
\begin{equation}
  P(\deltaE)=\frac{\exp\left[-\alpha^2\left\{(1+\delta)^{-1/\alpha}-1\right\}^2/(2\sigma_\psi^2)\right]}
  {(1+\deltaL)^{-2-1/\alpha}\left(1+\frac{\alpha-1}{2\alpha}\sigma_\psi^2\right)\sqrt{2\pi\sigma_\psi^2}}
  \label{eqn:pdfdelta}
\end{equation}
The middle factor in the denominator is $\avg{V}$, which we found to
have the form $1+V_2(\alpha)\sigma_\psi^2$ for $1<\alpha <3$.  The form for
$V_2$ gives the analytically calculable coefficient at
$V_2(\alpha=1,2,3)=(1, 1/4, 1/3)$, and matches the numerically estimated
coefficient at other $\alpha$'s, including the SC $V_2(3/2)=1/6$.

\subsection{Reduced non-Gaussianity in the log-density}
\begin{figure}
  \begin{center}
    \includegraphics[scale=0.8]{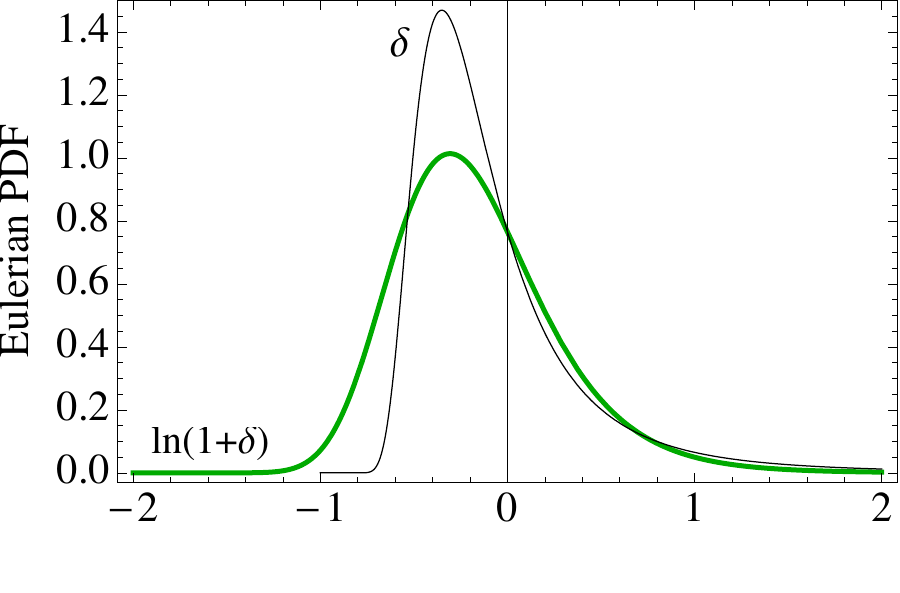}
  \end{center}  
  \caption{Eulerian PDFs of $\delta$ and $A=\ln(1+\delta)$ from
    Eqs.\ (\ref{eqn:pdfdelta}) and (\ref{eqn:pdfa}), setting $\alpha=3/2$
    and $\sigma_\psi=0.5$.}
  \label{fig:pdfalpha}
\end{figure}

\begin{figure}
  \begin{center}
    \includegraphics[scale=0.8]{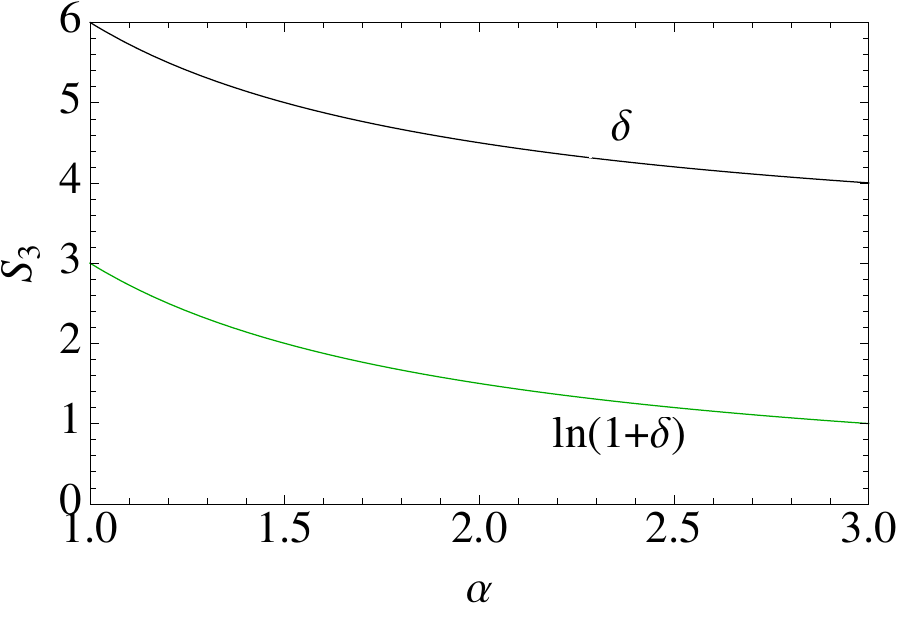}
  \end{center}  
  \caption{Skewness parameters $S_3$, for both the overdensity
    $\delta$ and the log-density $A$, letting $\sigma_\psi\to0$, as a
    function of $\alpha$, using Eqs.\ (\ref{eqn:pdfdelta}) and
    (\ref{eqn:pdfa}).  The curves are computed numerically, but match
    the relations in Eq.\ (\ref{eqn:s3deltaa}).}
  \label{fig:s3}
\end{figure}

\begin{figure}
  \begin{center}
    \includegraphics[scale=0.4]{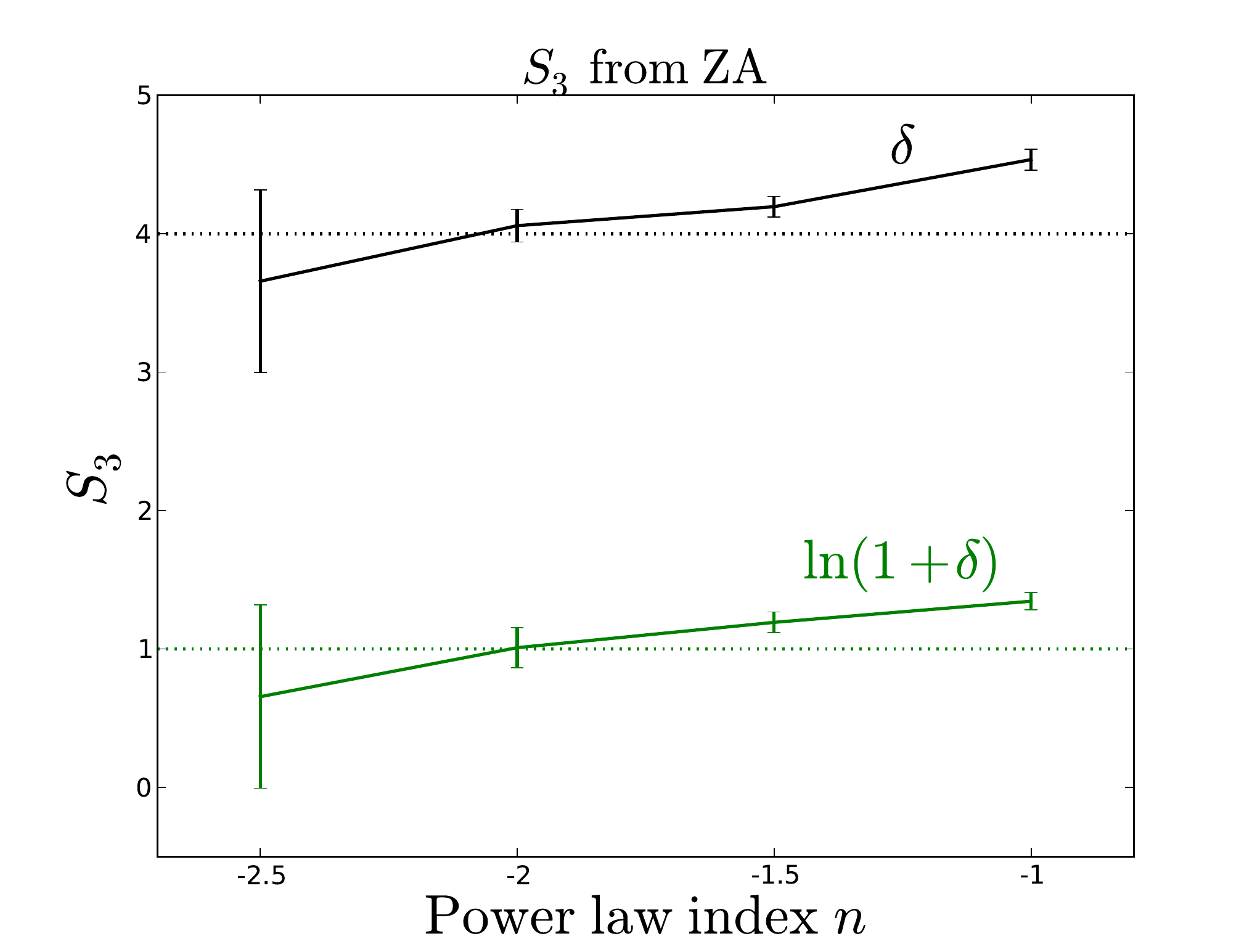}
  \end{center}  
  \caption{Volume-weighted skewnesses of $A$ and $\delta$, measured
    from Voronoi-cell volumes around particles in Zel'dovich
    realizations with power-law power spectra of indices $n$.  The
    dotted curves are the asymptotic $\sigma_\psi\to 0$ values of
    $S_3$ in the cubic-mass-element model.  }
  \label{fig:interparts3}
\end{figure}

As is well-known in cosmology \citep[\eg][]{colesjones,Colombi,nss09},
the PDF of the log-density, $A\equiv\ln(1+\delta)$, is much more
Gaussian than the PDF of $\delta$.  One way to understand this is that
$P(\delta)$, unlike $P(A)$, is tied down to zero at $\delta=-1$.
Because it is so easy to do in the $\alpha$-parameterized model, here
we give some explicit formulae for the PDF and skewness of $A$.

A Gaussian $\psi$ distribution transforms to
\begin{equation}
  P(\AL)=\frac{\exp\left[-\frac{A}{\alpha}-\frac{\alpha^2}{2 \sigma_\psi^2} \left(e^{-A/\alpha}-1\right)^2\right]}
{\sqrt{2\pi\sigma_\psi^2}}.
  \label{eqn:pdfaL}
\end{equation}
\begin{equation}
  P(\AEul)=\frac{\exp\left[-A(1+\frac{1}{\alpha})-\frac{\alpha^2}{2 \sigma_\psi ^2} \left(e^{-A/\alpha}-1\right)^2\right]}
  {\left(1 + \frac{\alpha-1}{2\alpha}\sigma_\psi^2\right)\sqrt{2\pi\sigma_\psi^2}},
  \label{eqn:pdfa}
\end{equation}
with the same $\avg{V}$ factor in the denominator as in Eq.\ (\ref{eqn:pdfdelta}).

Fig.\ \ref{fig:pdfalpha} shows Eulerian (volume-weighted) PDFs of
$\delta$ and $A$ using the SC $\alpha=3/2$, with $\sigma_\psi=0.5$.
Even at this modest $\sigma_\psi$, $A$ is visibly more Gaussian than
$\delta$.

The first-order non-Gaussianity statistic is the skewness $S_3=\langle
\delta^3\rangle/\langle\delta^2\rangle^2$, which has been worked out
perturbatively in the mildly non-linear regime, in both Eulerian
perturbation theory (EPT) and in the Zel'dovich approximation. Without any
smoothing, to Eulerian second (`tree') order, $S_3=34/7\approx4.86$ in an
Einstein-de Sitter (EdS) universe \citep{peebles}, with small
corrections in the \LCDM\ case.  In the Zel'dovich approximation,
$S_3^{\rm Zel}=4$ \citep{BernardeauSkewness1994,FryScherrer1994}, a
bit lower.  If the skewness measurement is done
smoothing over equal-sized Eulerian cells, a term is added that
depends on the (local) power-spectrum slope $n_{\rm
  eff}=d\ln\sigma^2(R)/d\ln R$, where $R$ is the smoothing radius.
With smoothing, $\gamma=-(n_{\rm eff}+3)$ is added to $S_3$.

In several cases, we found that $S_3$ in the limit of small
fluctuations was reduced by 3 when measuring it from $A$ instead of
$\delta$.  The simplest example is the exact lognormal distribution,
for which, analytically, the skewness of the log-density $S^A_3=0$,
and $S_3=3$, for any $\sigma_A$.  A more general reduction of the
skewness by 3 may only hold precisely in other cases in the limit
$\sigma_\psi\to0$.

Fig.\ \ref{fig:s3} shows our numerical (using Mathematica) estimate of
$S_3$ and $S_3^A$ as a function of $\alpha$, letting
$\sigma_\psi\to0$.  The relations
\begin{equation}
  S_3=3/\alpha+3, {\rm and}\ S_3^A=3/\alpha
  \label{eqn:s3deltaa}
\end{equation}
match the numerical solution, as well as our and PS97's analytical
findings (at $\alpha=1$, 3/2, and 3).  $S_3(\alpha=3/2)=5$ is close to
the full-gravity value from EPT mentioned above, $34/7\approx4.86$,
which is also the leading-order result in the full spherical-collapse
model in the EdS case
\citep{FosalbaGaztanaga1998I,BernardeauEtalReview2002}.  In fact,
$S_3=5$ in the limit $\Omega_\Lambda\to0$ of the full SC dynamics.
This match to the $\alpha=3/2$ skewness is not surprising since the
$\alpha=3/2$ model arises in the same limit.

It may be worth investigating tuning the $\alpha$ parameter further,
for example to investigate a model with $\alpha=21/13\approx1.62$,
which would exactly give the EdS EPT and SC skewness, and at the same
time give a critical collapse $\psil=-1.62$ for collapse, nearly the
full nonlinear spherical-collapse value of $\psil\approx -1.686$.

As PS97 found, in the cubic-mass-element approximation as
$\sigma_\psi\to 0$, $S_3=4$ as in the ZA, although $S_3$ diverges from
4 differently than in the ZA as $\sigma_\psi$ departs from 0.  We
numerically investigated $S_3$ in the ZA, measuring the
volume-weighted particle-density skewness parameter $S_3$ from several
ZA realizations.  The Zel'dovich-produced \LCDM\ initial conditions of
the simulation shown in subsequent sections have $S_3=4.01$, and
$S_3^A=1.07$.

Fig.\ \ref{fig:interparts3} shows the volume-averaged skewness in
$\delta$ and $A$ measured from the distribution of particle Voronoi
densities in Zel'dovich simulations with power-law power spectra.  The
error bars are the dispersions among 3 realizations at each $n$.  As
$n$ decreases, large-scale over small-scale fluctuations dominate.
$S_3$ diverges somewhat from 4 at high $n$. In the context of the
$\alpha$ model, it makes intuitive sense that the isotropic,
cubic-mass-element would be most valid for low $n$, where large-scale
fluctuations dominate.  As $n$ increases, mass elements cease to
expand or contract isotropically, so the effective $\alpha$ decreases.

Note that this measurement, although it is Eulerian (volume-weighted),
does not include a smoothing of the type that would add a $\gamma$
term to $S_3$, since there is no fixed cell size.  Of course, the
realizations have finite (mass) resolution; thus `no smoothing' is
not meant to imply infinite spatial and mass resolution.  A $\gamma$
term from a fixed Eulerian cell size would cause $S_3$ to depart from
4 in the opposite way than we observe when $n$ is increased from
-3. The $\sigma_\psi$ used to generate each realization was held fixed
at 0.02, and the power-law index $n$ was varied from $-1$ to $-2.5$.
Error bars show the dispersion from three different realizations
analyzed at each $n$.

From these numerical results, it appears that in the ZA, as well as in
the $\alpha$ approximation, a log transform reduces $S_3$ by 3.  It
would be interesting to show how widely this property holds, a
question for later work.

\section{Behavior of $\bpsi$ in full gravity}
\label{sec:fullgravity}

\begin{figure}
  \begin{center}
    \includegraphics[scale=0.425]{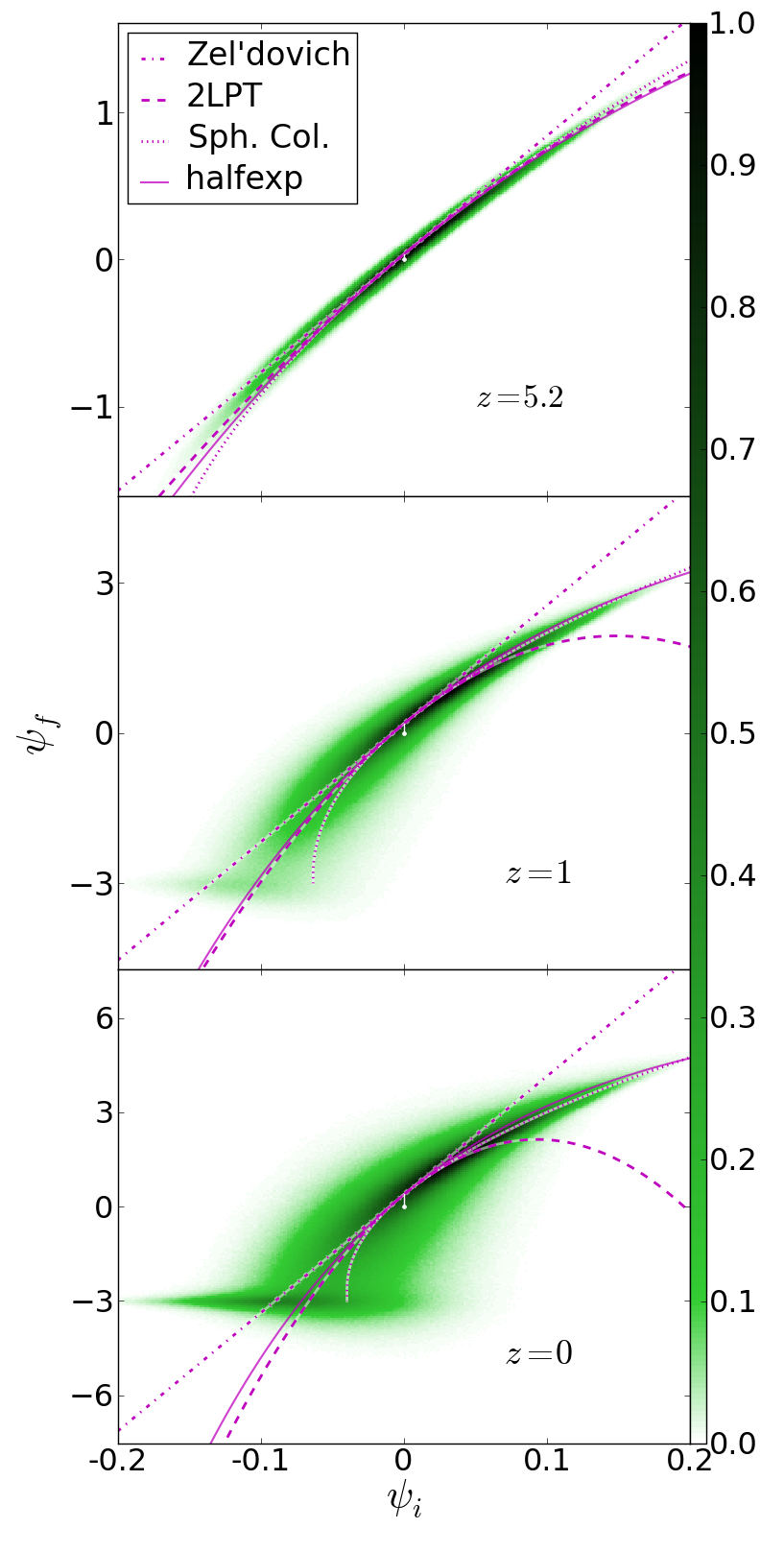}
  \end{center}  
  \caption{In green, two-dimensional histograms showing the relation
    between $\psi_i$ at the initial redshift of 49, and $\psi_f$,
    measured at the redshifts listed.  In a `local Lagrangian'
    approximation, $\psi_f$ is simply a function of $\psi_i$.  With
    time, the dispersion between the two grows, but even at $z=0$, the
    dispersion is rather low (considering the stretched color scale).
    Various local Lagrangian approximations are shown in magenta.  The
    2LPT curve, from Eq.\ \ref{eqn:psiparab}, is a lower bound on
    $\psitwo$.  For a high-density mass element, $\psi_f$ migrates
    downward until the element collapses, giving $\psi_f=-3$, about
    which it oscillates afterward.  The small white circles and lines
    about (0,0) show the magnitude of the shift in the approximation
    curves caused by enforcing $\avg{\psi_f}=0$.}
  \label{fig:divpsihists}
\end{figure}

\begin{figure*}
  \begin{minipage}{175mm}
    \begin{center}
      \includegraphics[scale=0.55]{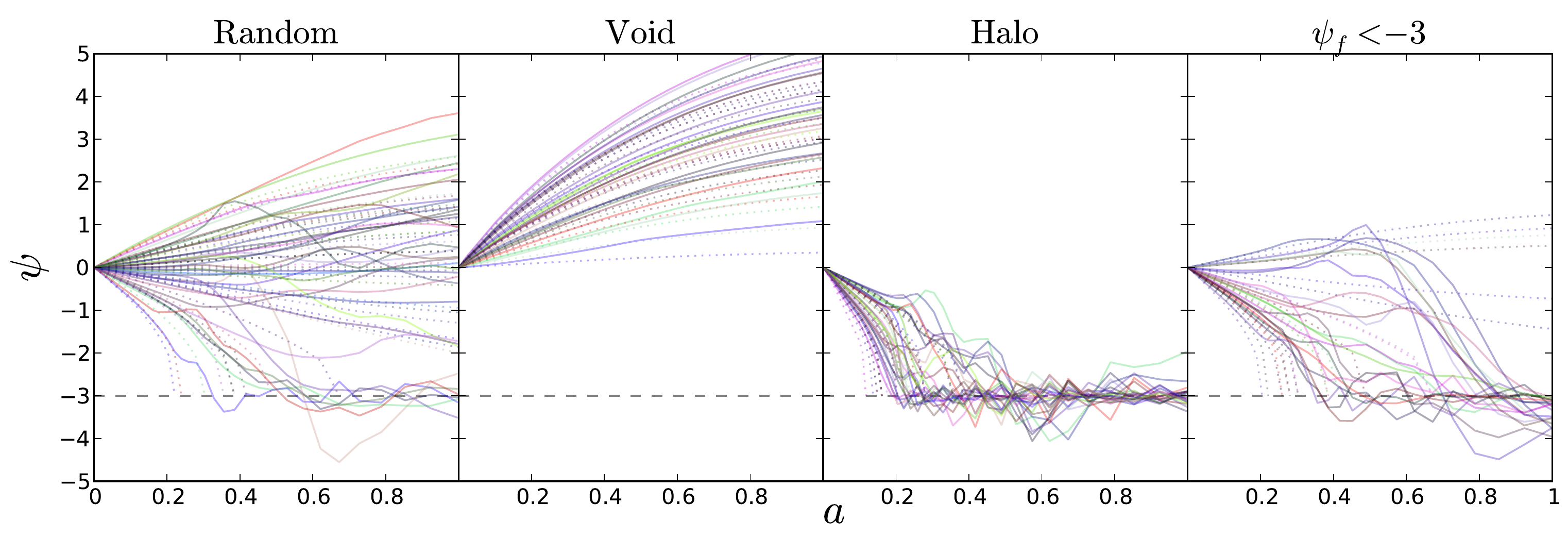}
    \end{center}  
    \caption{Trajectories in $\psi$, as a function of cosmic scale
      factor $a$, of four sets of particles, 25 in each panel, drawn
      from the two-dimensional slice shown below of an $N$-body
      simulation.  From left to right: (1) A random set of
      particles; (2) particles within a Lagrangian distance of
      2.3\hmpc\ of the maximum-$\psi_i$ (lowest-initial-density)
      particle in the slice; (3) particles within a Lagrangian
      distance of 2.3\hmpc\ of the minimum-$\psi_i$
      (highest-initial-density) particle in the slice; (4) a random
      subset of particles with $\psi_f<-3$.  Dotted curves show
      $\psisc(\psi_i)$, using Eq.\ (\ref{eqn:psisc}), colored the same
      as each particle's actual trajectory. The dashed line indicates
      the $\psi=-3$ collapse `barrier.'}
    \label{fig:psitraject}
  \end{minipage}
\end{figure*}

Here we compare these theoretical estimates to what actually occurs in
an $N$-body simulation.  The simulation has $256^3$ particles in a
200-\hmpc\ box, run with a vanilla \LCDM\ cosmology ($\Omega_m=0.3$,
$\Omega_\Lambda=0.7$, $\sigma_8=0.9$, $h=0.73$, $n_s=1$).  The initial
conditions were generated at redshift $z=49$ using the Zel'dovich
approximation, and run with the {\scshape Gadget 2} \citep{Gadget2}
code.  The spatial stretching parameter $\psi(\bq)\equiv\divqpsi(\bq)$
is measured by numerically differencing neighboring particle
positions.

Fig.\ \ref{fig:divpsihists} shows the evolution of $\psi(\bq)$ with
redshift, showing 2D histograms of initial to final $\psi$ at
different snapshots.  At moderate and low densities, the initially
straight line at $z=49$ grows bent, and develops a scatter
(accentuated somewhat by the color scale).  At high densities, there
is a critical value $\psi=-3$, signifying collapse of the mass
element.  In an idealized case where a Lagrangian patch of particles
contracts into a single point (a `halo'), $\psi=-3$.  Here, the
Lagrangian divergence of the particle-position field $\bx(\bq)$ is
zero; $\psi=\divqx-3=-3$.  

The locality of the $\psi_f$-$\psi_i$ relationship can be measured
with the dispersion in these 2D histograms.  At $z=0$, as expected,
the locality is high in void regions, and degrades at higher
densities, but, considering the stretched color scale, the
relationship is still rather tight.

As before, $\psi$ was measured by differencing the positions of
Lagrangian-neighbor particles.  The development of the $\psi=-3$ peak
in $\psi_f$ is sensitive to the method of measuring $\psi$; it does
not appear if the divergence is measured in Fourier space.  Perhaps
this arises from sharp edges being difficult to describe precisely in
Fourier space.

Once a particle crosses the $\psi=-3$ barrier, $\psi$ changes
stochastically, but stays around $-3$, since Lagrangian neighbors stay
nearby in a halo compared to the Lagrangian interparticle separation
(assuming somewhat low mass resolution).  Thus a collapsing mass
element's $\psi$ value evolves as though it were in a waterfall: it
descends with time as the mass element's density increases, then hits
a `surface' at $\psi=-3$, about which it then bobs around.

For $\psi_i<0$ but $\psi_f>-3$ (at high densities before actual
collapse), the best of the above approximations for $\psi_f>-3$ seems
to be $\psiparab$ (Eq.\ (\ref{eqn:psiparab})), lying between the
overpredicting Zel'dovich (Eq.\ (\ref{eqn:zeld})) and the
underpredicting SC [Eq. (\ref{eqn:psisc})] predictions.  For $\psi\ge0$
(at low densities), the SC prediction is best, again lying between the
two rather poor alternatives.  Curiously, these two approximations are
both parabolic, one in $\psi_i$, and the other in $\psi_f$.

We found empirically that another approximation shown,
\begin{equation}
  \psihalfexp=D^{1/2}\left(1-e^{-D^{1/2}\psi_i}\right),
  \label{eqn:halfexp}
\end{equation}
works well for both high and low $\psi_i$.  However, we caution that
to our knowledge it lacks theoretical motivation, and has strange
behavior at very high $\psi$ (higher than plotted here), asymptoting
to $D^{1/2}$.

The curves do not precisely go through the origin, which appears as a
white dot.  The offsets, shown by small white lines, ensure that
$\avg{\psisc}=0$.  For simplicity, we apply the same offset to all
curves.  The numerical value of this offset is similar for the various
approximations, except for the ZA, for which $\psi_f$ is always
symmetric about zero.  This $\avg{\psi}=0$ condition ensures that
there is no mean comoving expansion or contraction.

The SC approximation, $\psisc$, predicts a particular trajectory of
$\psi$ with time, depending only on the local $\psil$.  To investigate
how well this approximation holds with time, for particles in
different environments, in Fig.\ \ref{fig:psitraject} we show
trajectories in $\psi$ for particles in various classes: a random
selection of particles; particles near the highest-initial-density and
the lowest-initial-density particles; and particles with $\psi_f<-3$.
At early epochs ($a\lesssim0.2$), $\psi$ tracks $\psisc$
well. Subsequently, particles participating in nonlinear structures
can get seriously derailed (e.g., the rightmost panel).  Still, there
are many particles for which $\psi$ continues to track $\psisc$.  In
the deepest void, for example (second-left panel), the form of the
expansion roughly tracks the SC prediction, but is skewed upward,
perhaps due to the extremity of the void.

\begin{figure*}
  \begin{minipage}{175mm}
    \begin{center}
      \includegraphics[scale=0.6]{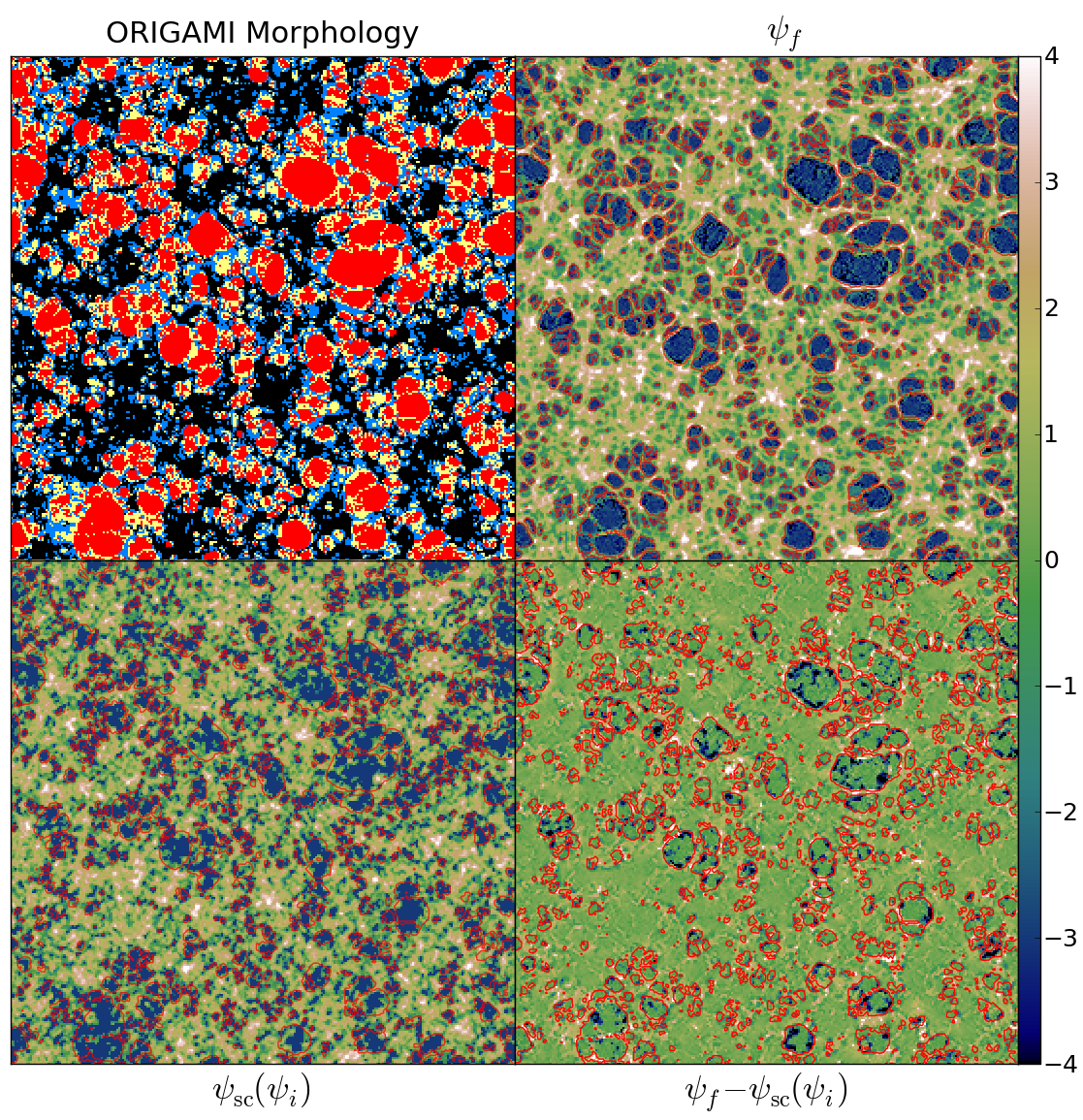}
    \end{center}  
    \caption{Quantities measured on a Lagrangian sheet of 256$^2$
      particles from a 256$^3$-particle simulation of box size
      200\hmpc, run to redshift 0. Each pixel corresponds to a
      particle.  Upper left: the \org\ morphology of the particle
      (void, wall, filament and halo particles are colored black,
      blue, yellow and red).  Upper right: the Lagrangian divergence
      of the displacement field, $\psi_f$.  The `lakes' are Lagrangian
      regions that have collapsed to form haloes.  Lower left: a
      prediction from the initial-conditions $\psi_i$, using the
      spherical-collapse formula (\ref{eqn:psisc}).  $\psi_f$ is set
      to -3 if $\psil<-3/2$.  Lower right: the difference between
      upper left and lower right.  The red contours mark the
      boundaries of haloes as identified by \org.  }
    \label{fig:comparemap}
  \end{minipage}
\end{figure*}

Fig.\ \ref{fig:comparemap} shows $\psi_f$, measured at $z=0$ for
$256^2$ particles occupying a flat Lagrangian sheet from this
simulation.  Some similar figures appear in \citet{MohayaeeEtal2006}.
The quantities are plotted in Lagrangian (initial-conditions)
coordinates, with each pixel corresponding to a particle on the square
lattice.  Also plotted are the \org\ \citep{FalckOrigami2012}
morphologies of the particles in the sheet, as well as the result of
applying the spherical-collapse equation (\ref{eqn:psisc}) to the
initial conditions.  A particle's morphology in the \org\ algorithm
measures the number of Lagrangian axes along which other particles
have crossed it in Eulerian space.  A halo particle has been crossed
by other particles along 3 orthogonal axes; filament, wall, and void
particles have been crossed along 2, 1, and 0 orthogonal axes.  In the
bottom panels, if $\psil<-3/2$, Eq.\ (\ref{eqn:psisc}) has no
solution, i.e.\ the mass element has collapsed; in this case, we set
$\psisc=-3$.  Contours indicate the boundaries of \org\ halo regions.

The color scheme suggests a topographical analogy, when working in
Lagrangian coordinates: as time passes, $\psi$ departs from zero, in a
way largely prescribed by its initial value.  However, in overdense
regions where it is decreasing, it is not allowed to plummet
arbitrarily; where collapses occur, `lakes' form, where
$\psi$ becomes $\approx-3$.

In the upper-right panel, the blue `lakes' of $\psi_f$ correspond
quite well to halo regions as identified by \org.  A simple
halo-finder comes to mind, connecting particles on the Lagrangian
lattice with $\psi$ under some threshold, approximately $-3$.  We did
try a simple implementation of this, but had difficulty finding a
simple threshold to characterize all haloes, since there are roughly
as many halo particles with $\psi>-3$ as $\psi<-3$.  Still, we suspect
a halo-finder along these lines could be quite successful.

The bottom panels compare $\psisc$ and $\psi_f$.  As in
Fig.\ \ref{fig:divpsihists}, the two match quite well in void regions,
but in high-density regions, the correspondence is rougher.  The rms
difference between $\psisc$ and $\psi_f$ in this simulation is 1.31.
The agreement in overdense regions can be improved by using $\psisc$
for $\psi>0$, but $\psiparab$ for $\psi < 0$.  This reduces the rms
difference to 1.24.  These are to be compared to the standard
deviation of $\psi_f$ (i.e.\ the rms if it is approximated with its
mean, 0), which is 2.0.

Another interesting feature of this plot is that
$\psi_f-\psisc(\psi_i)$ often plummets (i.e.\ becomes large and
negative) on the Lagrangian outskirts of haloes.  There are a couple
of possible reasons for this.  The particles have been dragged into
the halo at late times, and may not be overdense initially.  Also, as
they have just fallen into haloes, their cubic Lagrangian volumes have
likely just been swapped. Thus when a particle first collapses, $\psi$
generally overshoots $-3$, as also shown in
Fig.\ \ref{fig:psitraject}.

\begin{figure}
  \begin{center}
    \includegraphics[scale=0.5]{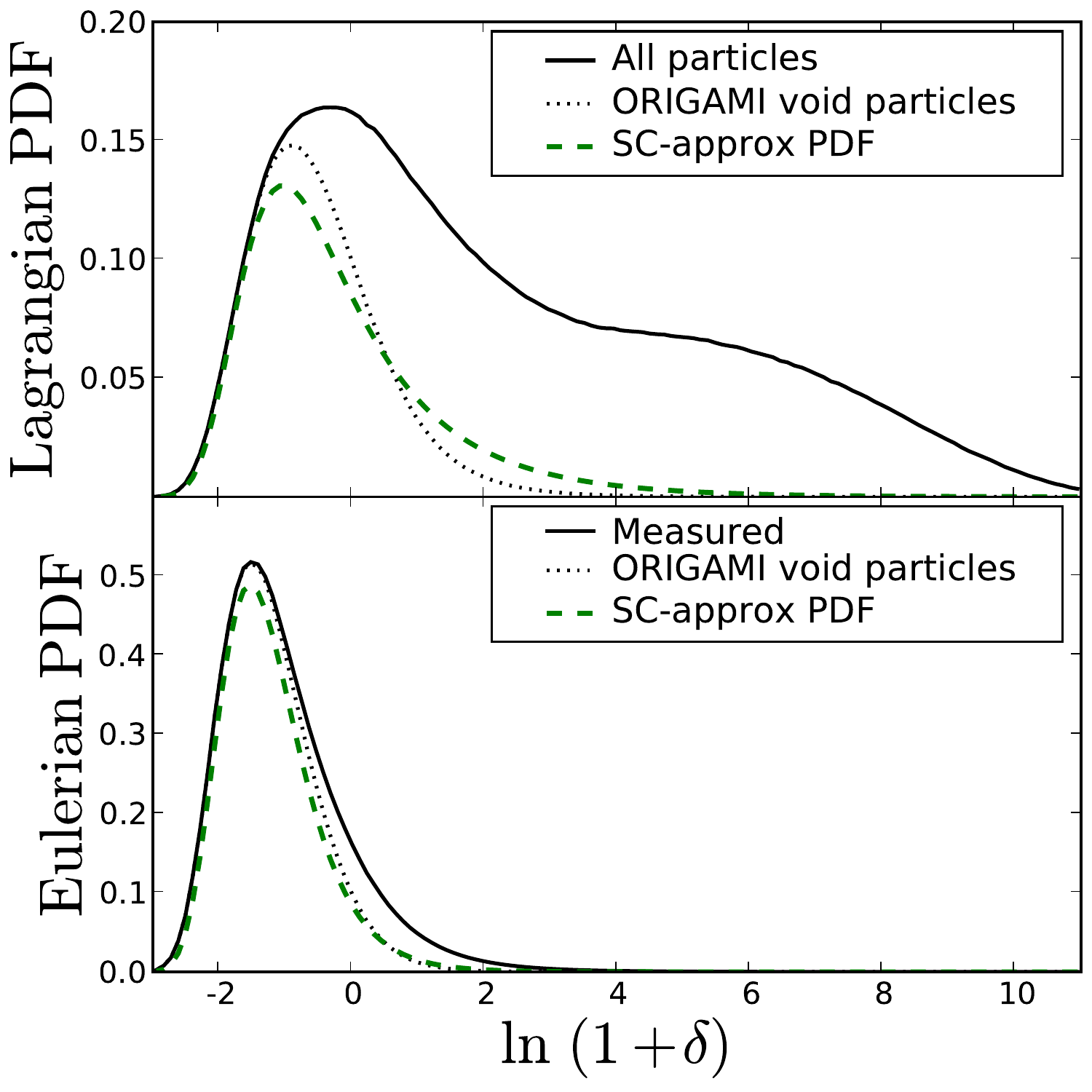}
  \end{center}  
  \caption{Voronoi-measured histograms of $A=\ln(1+\delta)$, including
    all particles, and also including only \org-identified void
    particles, which have undergone no stream-crossing.  The
    theoretical PDFs are as in Eqs.\ (\ref{eqn:pdfaL}) and
    (\ref{eqn:pdfa}), applying a normalization correction equal to the
    fraction of \org\ void particles in the simulation.  The top
    histogram is `Lagrangian' as in mass-weighted, with each particle
    contributing equally.  The bottom, Eulerian histograms are also
    measured using the Voronoi tessellation; they are simply the
    `Lagrangian' PDFs multiplied by $V=1/(1+\delta)$.}
  \label{fig:pdfs}
\end{figure}

Fig.\ \ref{fig:pdfs} compares Voronoi-measured, mass-weighted PDFs to those
assuming that each volume element evolves independently, i.e.\ the
PDFs of Eqs.\ (\ref{eqn:pdfdeltaL}-\ref{eqn:pdfa}).  Again, the
`Lagrangian' PDF is not truly Lagrangian, since the density estimate
for a particle includes all other particles, not just its Lagrangian
neighbors.  Thus stream-crossing boosts each particle's density
according to the locally overlapping number of streams, populating the high-density `shelf' that poorly matches the approximation.

At low densities, the approximations match the measurements well.
PS97 also found this, albeit in simulations without as much structure.
However, a normalization correction was necessary for the agreement in
Fig.\ \ref{fig:pdfs}.  A greater fraction of particles than predicted
by the SC approximation leave the SC tracks to populate the
high-density tail, as in Fig.\ \ref{fig:psitraject}.

If the SC approximation were precisely accurate in describing both the
density evolution up to stream crossing, and the fraction of particles
whose Lagrangian volumes have collapsed, no normalization correction
would be necessary in Eq.\ (\ref{eqn:pdfaL}), since as noted earlier,
we did not divide by the integral over the PDF to assure a PDF
integrating to unity.  At this value of $\sigma_\psi$,
Eq.\ (\ref{eqn:pdfaL}) integrates to 0.71, which can be calculated
with a simple Erf expression giving the fraction of particles with
$(1+\frac{2}{3}\psil)<0$ (the critical $-\psil=1.5$, intriguingly near
the critical density for collapse, 1.69).

Instead, we found that a smaller factor, 0.34, gave a good fit to the
low-density tail.  This matches the fraction of void particles in this
simulation snapshot as measured by the \org\ \citep{FalckOrigami2012}
algorithm.  An \org\ void particle has not been crossed by any other
particle over the course of the simulation.  We also show a curve
showing the PDF of only void particles.  The shape of this curve does
not quite match that from the SC-approximation PDF; that is, even
higher-density void particles are scattered a bit to higher densities.

The agreement between all curves looks much better in the bottom,
Eulerian panel.  This is because the PDF is simply the PDF in the
upper panel multiplied by the volume factor $V/\avg{V}$.  In terms of
the $x$-coordinate $A$, this is simply an exponential damping,
$e^{-A}$, bringing up the left side of the curve, and suppressing the
right side.

If the amplitude of this normalization correction can be estimated or
calibrated accurately, Eqs.\ (\ref{eqn:pdfaL}-\ref{eqn:pdfa}) seem to
provide a convenient estimate for the nonlinear density PDF, if a PDF
lacking the true, more-populated high-density tail is adequate.

\section{Particle realizations from the SC approximation}
\label{sec:realizations}

\begin{figure*}
  \begin{minipage}{175mm}
    \begin{center}
      \includegraphics[scale=0.6]{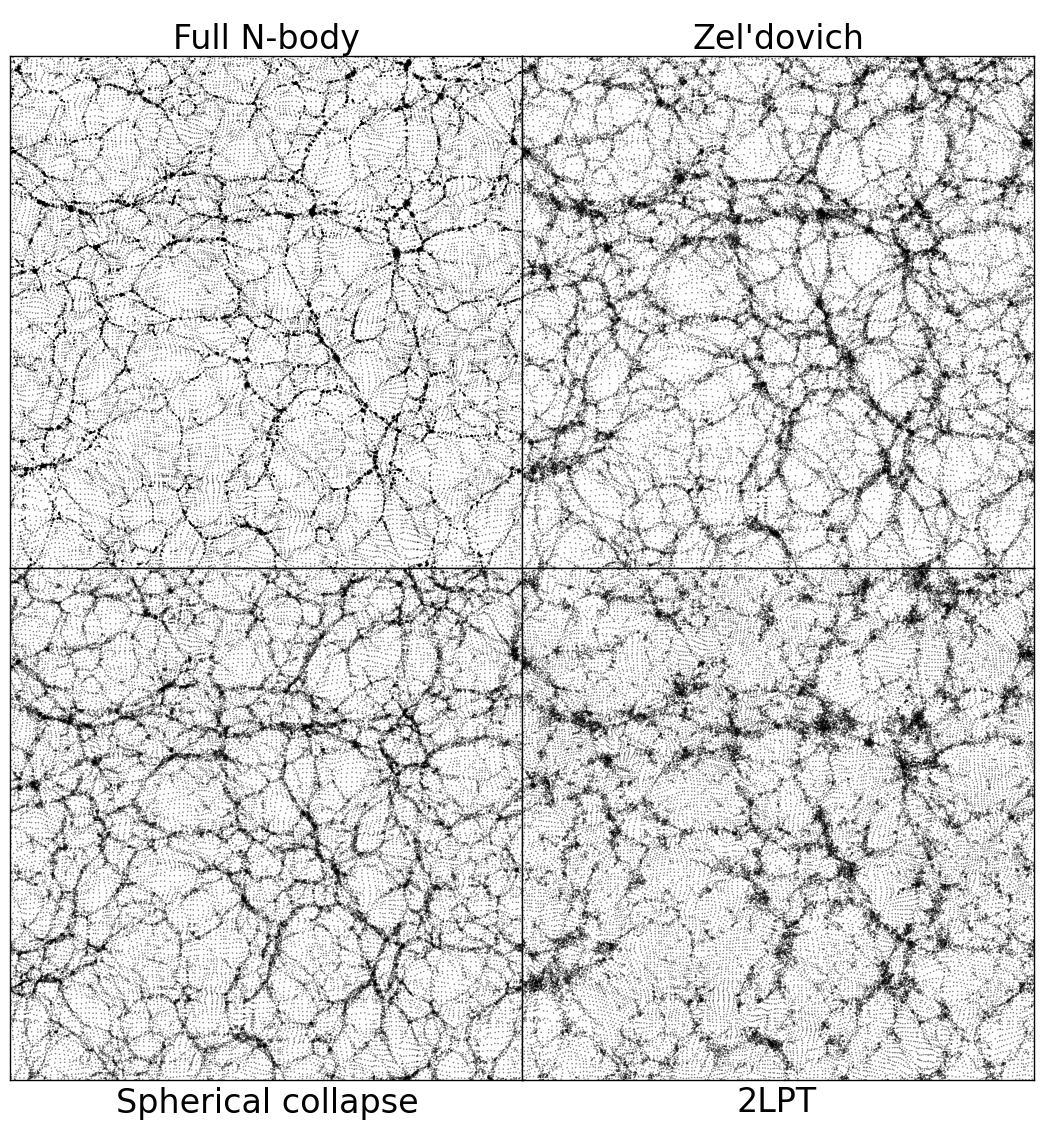}
    \end{center}  
    \caption{Redshift-zero Eulerian locations of particles occupying a
      256$^2$ sheet of a 256$^3$-particle \LCDM\ realization,
      projecting out the third dimension.  Clockwise from upper left,
      particle positions are determined using: a full $N$-body
      simulation; the Zel'dovich approximation; 2LPT; and the SC
      approximation (\ref{eqn:psisc}).  By eye, the SC approximation
      gives the results closest to full gravity.  }
    \label{fig:euler}
  \end{minipage}
\end{figure*}

Given that $\psisc$ tracks the evolution of $\psi$ well, we explored
how well it would work explicitly to advance $\psi_i$ to $\psi_f$
using Eq.\ (\ref{eqn:psisc}), additionally setting $\psi_f=-3$ for
collapsed particles where $\psil<-1.5$ (the singularity in
Eq.\ (\ref{eqn:psisc})).  This is a simple replacement for a
$\psi_i$-$\psi_f$ relationship in a Zel'dovich code, entailing only a
fast additional step (with a bit less computational effort than 2LPT).
This may be about as well as one can do with a local prescription
giving $\psi_f$ as a function of $\psil$, with no dependence on its
derivatives.  (For comparison to a 2LPT prescription, see Appendix D2
of \citet{Scoccimarro1998}).  Here are all of the steps in our SC
procedure:
\begin{enumerate}
\item Generate a Gaussian random field from a linear power spectrum.
  This becomes $\psil$.
\item For $\psil\le-1.5$, set $\psi_f=-3$.  For $\psi_f>1.5$, set
  $\psi_f=\psisc+C$, where $\psisc$ is from Eq.\ (\ref{eqn:psisc}),
  and $C$ is a constant ensuring that $\avg{\psi_f}=0$ (easily
  measurable by summing up $\psi_f$ with $C=0$).  $C$ is typically
  small; for example, in Fig.\ \ref{fig:divpsihists}, $C$ is the length
  of the small white line attached to the white dot.
\item Take the inverse divergence of $\psi_f$ (in Fourier space,
  inverting Eq.\ (\ref{eqn:fourierdiv})) to get the displacement field
  $\bPsi$.
\end{enumerate}

Fig.\ \ref{fig:euler} shows the particle positions resulting from
advancing the initial ($z=49$) conditions of this simulation to $z=0$,
compared to its actual $z=0$ particle positions.  The SC approximation
gives a particle arrangement more visually similar to that using full
$N$-body dynamics than using either of the LPT relations, for example
producing more concentrated `haloes' where $\psi=-3$.  Still, they are
not as pointlike as we had hoped, perhaps related to the difficulty of
capture sharp edges in Fourier space.  For example, measuring the
divergence in Fourier space does not capture the $\psi_f=-3$ peak in
Fig.\ \ref{fig:divpsihists}.  A completely real-space
inverse-divergence algorithm might be more adept at producing tight
haloes, but it is not obvious how such an algorithm would work.

\begin{figure}
  \begin{center}
    \includegraphics[scale=0.4]{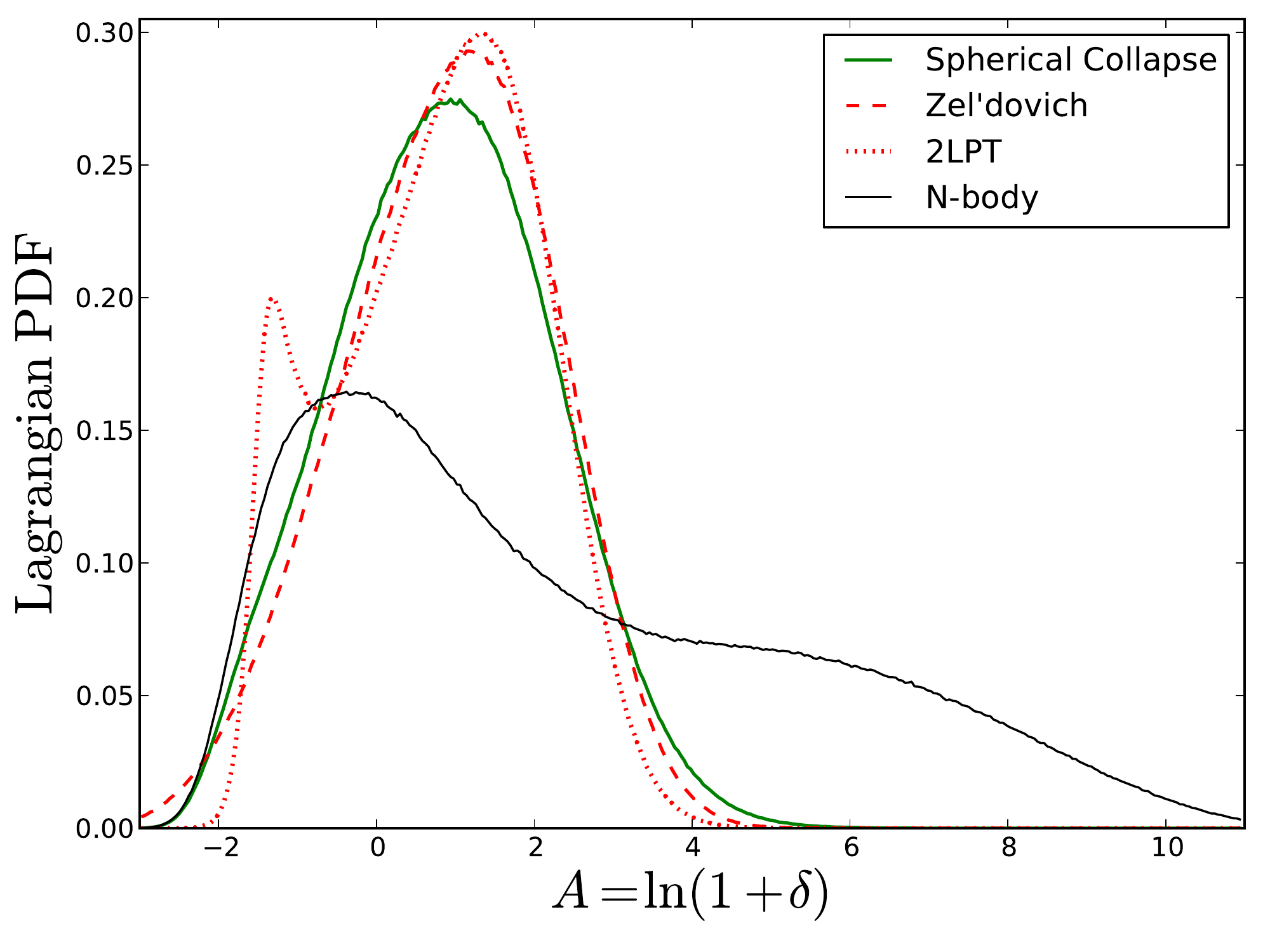}
  \end{center}  
  \caption{Mass-weighted histograms of Voronoi-estimated particle
    densities, for both the fully evolved initial conditions, and the
    three particle realizations shown in Fig.\ \ref{fig:euler}.  The
    SC approximation performs best at low densities; all
    approximations fail to capture the second peak or shelf of
    high-density halo particles present in the full simulation.  2LPT
    even produces a PDF peak at low densities.}
  \label{fig:vtfepdf}
\end{figure}

Fig.\ \ref{fig:vtfepdf} shows mass-weighted 1-point PDFs of each
particle distribution.  Here, the SC approximation gives the best
approximation to the true dynamics, especially for the lowest
densities, where the agreement is also quite good in
Fig.\ \ref{fig:divpsihists}.  There is a `shelf' of high-density
particles; these correspond to particles in haloes.  At higher
resolution, this locus becomes a peak \citep{FalckOrigami2012}.  The
low-density peak in the 2LPT PDF is a sign of its inaccuracy at this
rather high $\psil$ dispersion, $\sigma(\psil)=2.7$.  

\begin{figure}
  \begin{center}
    \includegraphics[scale=0.4]{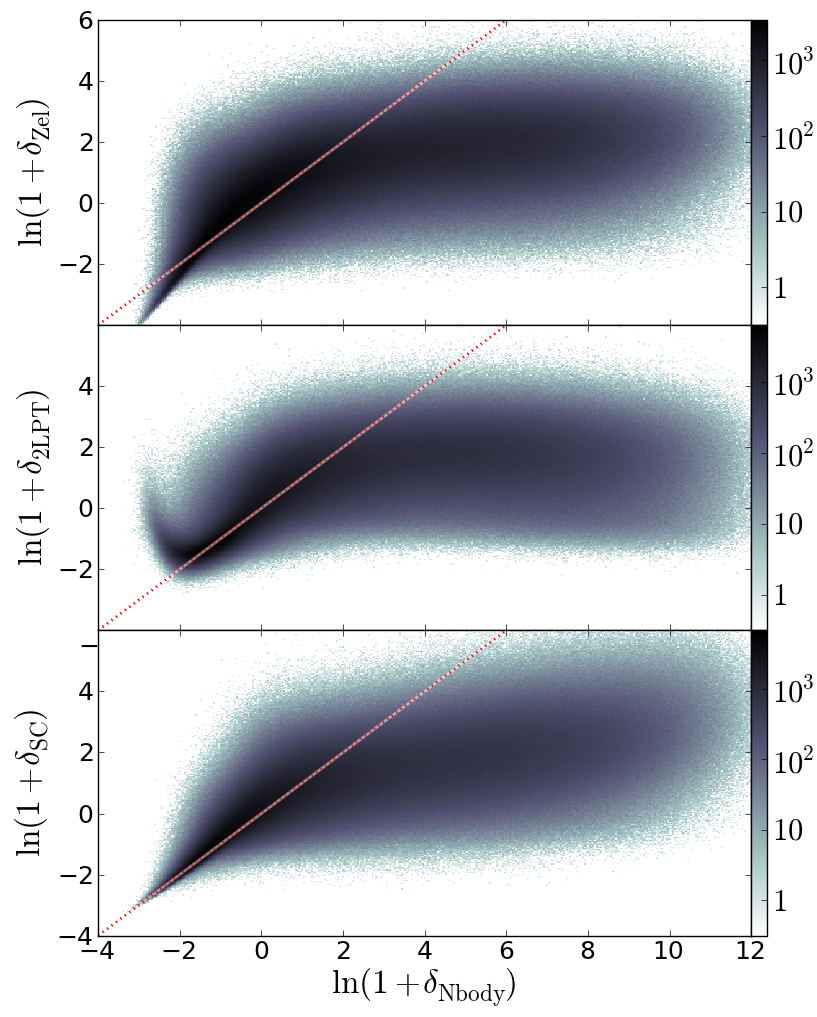}
  \end{center}  
  \caption{Two-dimensional histograms comparing particle densities
    evolved with full gravity (the $x$-axis) to densities in the
    approximately evolved particle distributions.  The dotted red
    lines show the ideal $y=x$ locus.  The SC-approximation-evolved
    particle distribution performs best.  Note the turn-up at low
    densities in 2LPT, where some overdense particles are predicted
    among particles that should be in the deepest voids.}
  \label{fig:denscomp}
\end{figure}

Fig.\ \ref{fig:denscomp} shows the same information in two-dimensional
histograms, comparing the full-gravity ($N$-body) $\delta$ to $\delta$
in the approximately evolved realizations.  Here, again $\psisc$
performs best.  In the low density region of the 2LPT scatter plot,
there are in fact overdense particles (reaching at maximum
$\delta\approx 8$), which should be extremely underdense in the final
conditions.  These middling overdensities are unlikely to produce
spurious haloes detected in a 2LPT realization, but there remains some
chance of that.

What are we to conclude about the reliability of 2LPT at low redshifts
for mock galaxy catalogs?  The work here is hardly an exhaustive
study, as it considers just the single simulation analyzed here.  But
for this simulation at $z=0$, the population of overdense particles
that should be underdense starts to be a worry.  This problem would be
even more severe if $\sigma(\psil)$ were increased, populating the
high-$\psi_i$ branch of the $\psiparab$ parabola.  One way to increase
$\sigma(\psil)$ is by increasing the mass resolution (since
fluctuations grow on small scales in a \LCDM\ universe), so we
recommend caution in using 2LPT realization at high resolution and low
redshift.  This is not surprising, of course; for high LPT accuracy,
$\sigma(\psil)$ should be $\lesssim 1$.  Fortunately, to our knowledge,
low-redshift uses of 2LPT have been at lower mass resolution than
this, resulting in an appropriately low $\sigma(\psil)$.

\begin{figure*}
  \begin{minipage}{175mm}
    \begin{center}
      \includegraphics[scale=0.6]{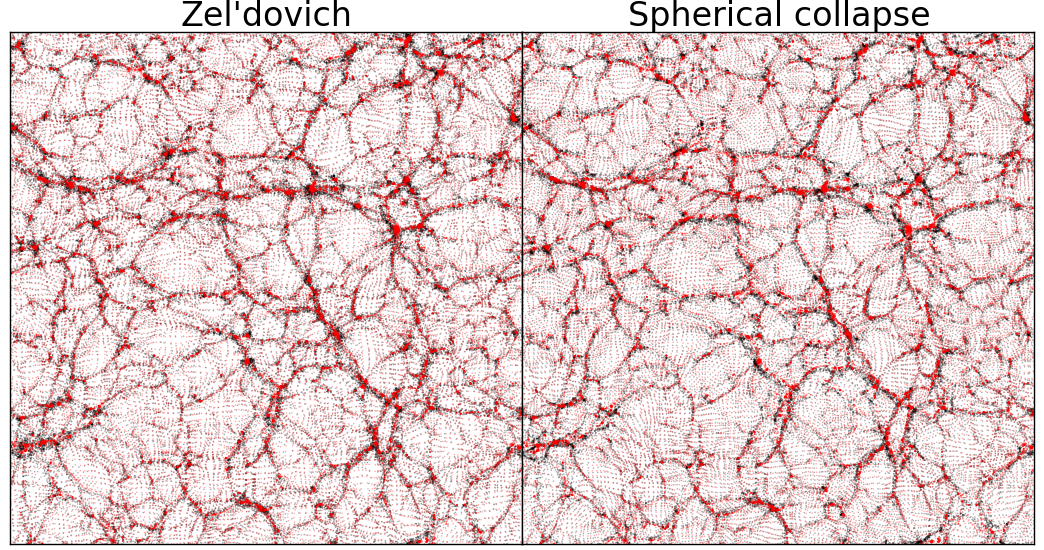}
    \end{center}  
    \caption{The Zel'dovich and SC panels of Fig.\ \ref{fig:euler},
      with the full $N$-body results overplotted in red.  While
      Zel'dovich gives artificially empty voids and fuzzier haloes, it
      gives somewhat more accurate large-scale flows than does SC.  }
    \label{fig:euler_overlay}
  \end{minipage}
\end{figure*}

While the SC approximation excels at predicting 1-point statistics and
a visually plausible particle distribution, unfortunately it seems to
have deficiencies, as well.  The SC approximation shifts the locations
of nonlinear structures more than does the ZA.  This is difficult to
see in Fig.\ \ref{fig:euler}, so we overplot the $N$-body and
approximate realizations in Fig.\ \ref{fig:euler_overlay}.  In 2LPT,
structures have similar locations as in the ZA.  We suspect that the
discrepancy in the SC case is largely from voids that collapse in the
full $N$-body case.  In the LPT approximations, overdense structures
surrounding the doomed voids collapse as they should, but this is
suppressed in the SC case.

The differences in large-scale flows in the SC show up in
root-mean-square errors in particle positions.  The rms errors of
particle positions compared to the full $N$-body dynamics using the
three approximations are, respectively for Zel'dovich, 2LPT, and SC,
1.61, 1.65, and 2.17\hmpc.

\begin{figure}
  \begin{center}
    \includegraphics[scale=0.4]{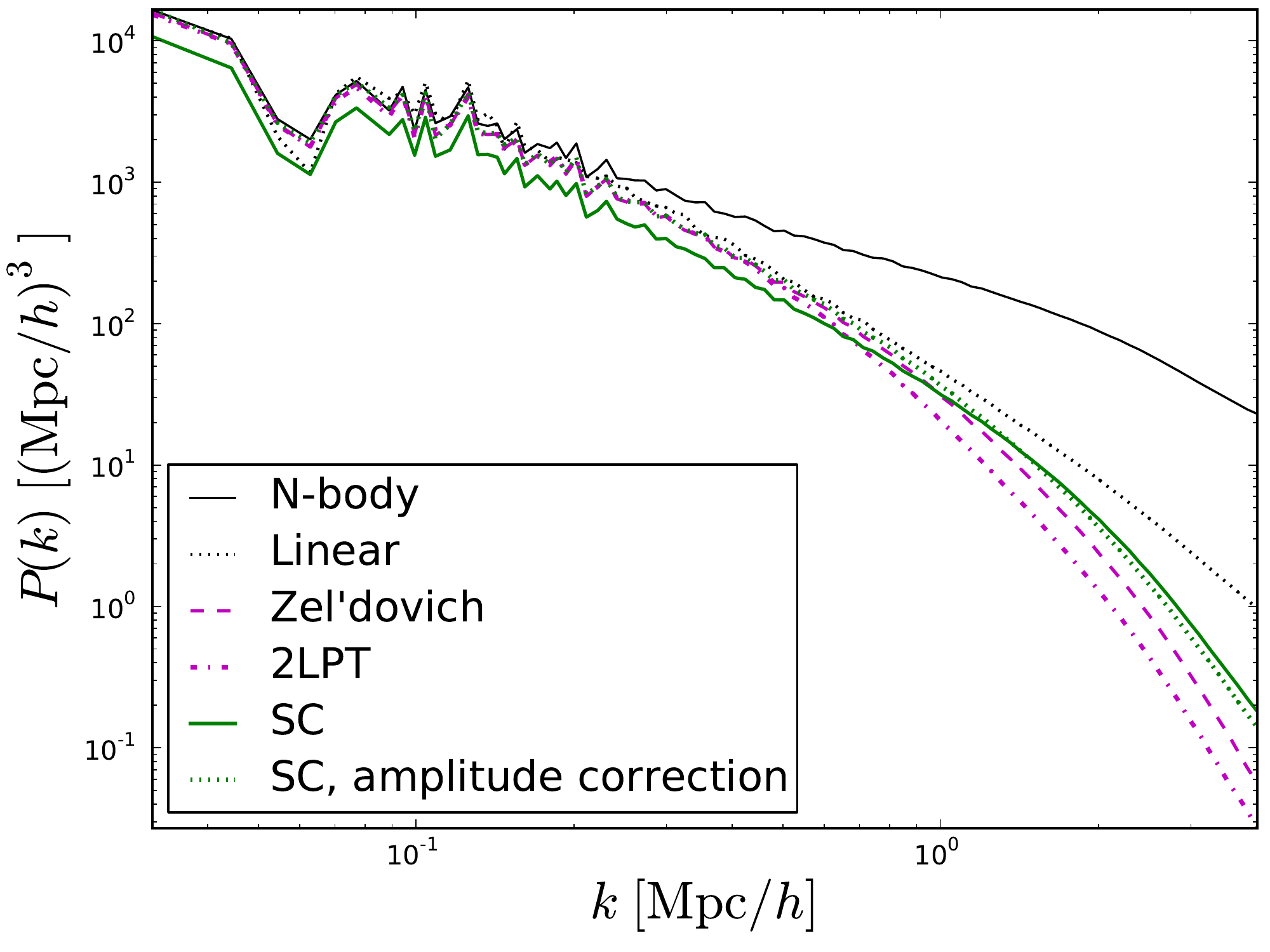}
  \end{center}  
  \caption{Matter power spectra in a 200-\hmpc\ $N$-body simulation at
    $z=0$, compared to power spectra of particle distributions
    displaced according to various approximations.}

  \label{fig:approxpowers}
\end{figure}

The LPT approaches are also more successful than the SC approximation
in predicting the low-redshift dark-matter power spectrum amplitude on large
scales, as shown in Fig.\ \ref{fig:approxpowers}.  The power spectrum
of particles displaced according to the SC approximation gives a
multiplicative bias on large scales of about 0.65 in this simulation,
although the SC power spectrum's shape is a bit closer to the shape of the
full nonlinear power spectrum, turning down at smaller scales than do
the LPT power spectra.

To measure the power spectra in Fig.\ \ref{fig:approxpowers},
particles were displaced according to each approximation, and then
assigned to cells on a 256$^3$ mesh using Nearest Grid Point mass
assignment.  Power spectra were then measured from these meshes,
correcting for shot noise.

It is possible to fix this large-scale normalization issue by
multiplying $\psil$ by a factor in Eq.\ (\ref{eqn:psisc}). In
Fig.\ \ref{fig:approxpowers}, scaling $\psil$ by an extra factor of
2.0 achieves this.  The factor was found by iteratively changing the
effective growth factor until the power spectra agreed on large
scales.  This `corrected' approximation also improves the agreement
between the SC and $N$-body-evolved particles, bringing the rms error
in particle positions down to 1.65\hmpc, at the level of the LPT
approximations.  However, undesirably, the `correction' also reduces
the SC power spectrum at small scales.

\begin{figure}
  \begin{center}
    \includegraphics[scale=0.4]{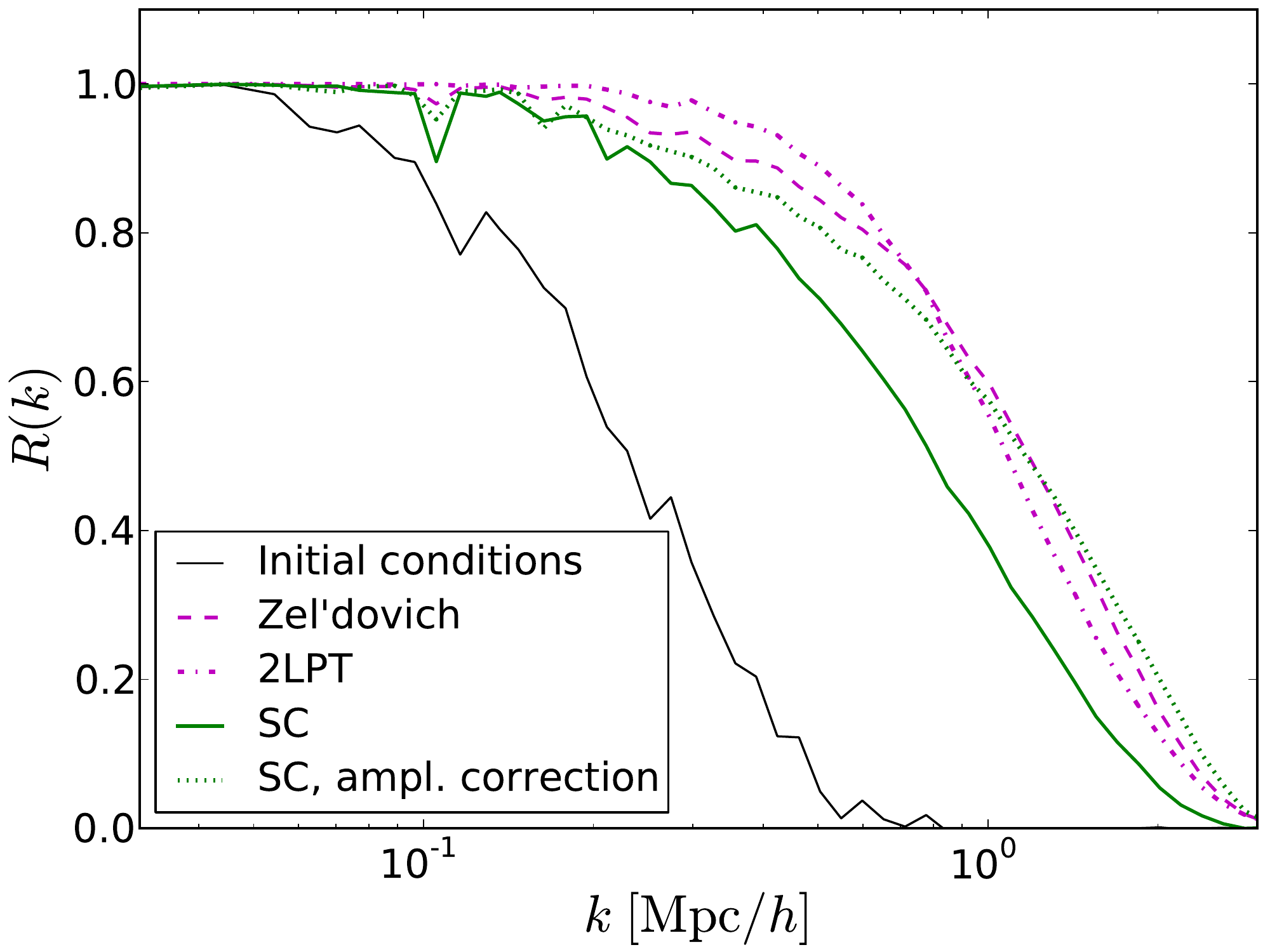}
  \end{center}  
  \caption{Fourier-space cross-correlation coefficients between the
    various approximately-evolved density fields and the particle
    distribution as evolved in the full $N$-body simulation.  The
    solid black line is essentially the non-linear propagator between
    the initial and final states; the Lagrangian cross-correlations
    are higher, indicating higher accuracy.}
  \label{fig:crosspowers}
\end{figure}

In Fig.\ \ref{fig:crosspowers}, we show the Fourier-space
cross-correlation coefficient $R(k) =
P_{\delta\times\delta^\prime}/\sqrt{P_\delta P_{\delta^\prime}}$ between
the $z=0$ density field in this $N$-body simulation, and several other
density fields.  The `initial conditions' curve is essentially the
propagator \citep[e.g.\ ][]{crocce}.  As pointed out recently by
\citet{TassevZaldarriaga2012}, the cross-correlation between
Zel'dovich and full gravitational dynamics is significantly higher at
small $k$ than the cross-correlation between Eulerian linear PT and
the full dynamics.  The agreement is even a bit better in 2LPT.  The
SC realization has poorer performance here, on the other hand,
although with the bias correction, $R(k)$ is the highest among the
approximations at large $k$.

Other ideas we tried in obtaining $\psi_f$ locally from $\psi_i$ were
to use $\psisc$ for $\psi_i > 0$, but $\psiparab$ for $\psi_i < 0$,
and also using $\psihalfexp$, based on the agreement in both cases to
the data in Fig.\ \ref{fig:divpsihists}.  These results were quite
similar to the simple $\psisc$ approximation, though.

It is quite likely that we could find some tweaking of these
approximations that would empirically give the best results.  However,
we did not explore this avenue exhaustively, since this best agreement
could be limited to this single simulation.  Still, it seems likely
that further optimization of the $\psi_i\to\psi_f$ mapping would be
fruitful.

\section{Summary and Conclusion}
\label{sec:conclusion}
In this paper, we examine $\psi$, the Lagrangian divergence of the
displacement field, arguably the most natural variable to quantify
large-scale structure in a Lagrangian approach.  The main results of
the paper are as follows:

\begin{itemize}
\item Even slight distortions of the initially uniform mesh of
  particles, quantified by $\psi$, produce a density distribution more
  lognormal than Gaussian.  It seems that for a wide class of models,
  the skewness parameter $S_3$ of the log-density field is reduced by
  3 compared to the skewness of the overdensity.

\item In 2LPT, the mapping from initial to final $\psi$ is roughly
  parabolic, allowing overdensities to form where there should be deep
  voids.  This does not seem to be a significant worry for a
  moderately low-density redshift-zero 2LPT realization ($\lesssim 1$
  particle per (\hmpcnosp)$^3$), but caution is recommended at higher
  resolution.

\item The spherical-collapse-fit formula (\ref{eqn:psisc}) describes
  the evolution of $\psi$ from initial conditions better than first-
  or second-order Lagrangian perturbation theory (LPT), up to halo
  formation.  This also allows for an approximation to the 1-point PDF
  of the density that works quite well for low-density, undisturbed
  (void) particles.

\item In LPT, $\psi$ gets arbitrarily small, indicating extreme stream
  crossing. In full gravity, however $\psi$ gets stuck around $-3$,
  signalling halo formation.  This is the value it would have if
  Lagrangian regions contracted exactly to pointlike haloes.

\item This knowledge of how $\psi$ evolves allows for a new method to
  produce final-conditions particle positions, based on this SC
  expression. Compared to LPT realizations, such SC realizations give
  reduced stream-crossing, and better visual and 1-point-PDF
  correspondence to the results of full gravity.  LPT realizations, on
  the other hand, give more accurate large-scale flows and large-scale
  power spectra, as well as improved cross-correlation to the density
  field evolved with full gravity. An empirical correction may be
  added to the SC formula that seems to fix some of these issues,
  however.

\end{itemize}

Our results suggest several possibilities for future work.  We did not
carefully investigate the new SC method of generating final-conditions
particle positions with respect to redshift and resolution.  We
suspect that the SC method could provide a good method of producing
relatively low-redshift initial conditions for simulations, if such a
thing is desired.  It is true that the large-scale power-spectrum bias
in this method is troubling, as are the shifts in large-scale flows,
but these could be tied to $\psi=-3$ collapses, and could be absent at
high redshift without stream-crossing.

The `barrier' at $\psi=-3$ could be useful for halo-finding in
$N$-body simulations.  Unfortunately, it seems not straightforward to
use this barrier to halo-find in a single snapshot of a simulation,
but other possibilities exist.  For instance, $\psi$ could be measured
at each timestep; if a particle ever has $\psi\le-3$, it could be
tagged as a halo particle.

It is also quite interesting to consider ways of predicting where
$\psi=-3$ from the initial conditions.  Such considerations may even
allow provide analytical mass functions.  Indeed, similar ideas have
been proposed using LPT \citep[e.g.\ ][]{MonacoEtal2002}. Another
possible approach may be to infer Lagrangian halo boundaries from
$\psisc(\psil)$ formula, as in the lower-left panel of
Fig.\ \ref{fig:comparemap}.  The true halo contours are often smoothed
versions of these contours, and perhaps could be obtained by a
combination of mathematical morphology techniques such as dilation and
erosion \citep[e.g.\ ][]{Serra1983} in Lagrangian space, as can be
useful in cleaning detected void boundaries \citep{PlatenEtal2007}.

In conclusion, $\psi$, a natural density-like variable in a Lagrangian
viewpoint, seems to be a rather useful quantity, with some extra
information that is not in the density itself.  It is fortunate that a
simple formula gives $\psi$'s behavior in voids, where dark energy is
most energetically dominant (if indeed it is a substance).  To
understand dark energy, understanding the stretching of the Lagrangian
mesh in voids is likely particularly important.

\section*{Acknowledgments}
I thank Guilhem Lavaux, Xin Wang, Alex Szalay, Istv\'{a}n Szapudi,
Miguel Arag\'{o}n-Calvo, Donghui Jeong and Nuala McCullagh for many
helpful discussions; Miguel Arag\'{o}n-Calvo additionally for use of
the simulation analyzed here; and an anonymous referee for a helpful
report.  I am grateful for financial support from the Gordon and Betty
Moore foundation and NSF award OIA-1124403.

\bibliographystyle{hapj}
\bibliography{refs}

\end{document}